\documentclass[conference]{IEEEtran}
\IEEEoverridecommandlockouts
\pdfoutput=1
\usepackage{cite}
\usepackage{amsmath,amssymb,amsfonts}
\usepackage{algorithmic}
\usepackage{graphicx}
\usepackage{textcomp}
\usepackage{xcolor}
\usepackage{subfig}
\usepackage{multirow}
\usepackage{stfloats}
\usepackage{siunitx}
\usepackage{float}
\def\BibTeX{{\rm B\kern-.05em{\sc i\kern-.025em b}\kern-.08em
T\kern-.1667em\lower.7ex\hbox{E}\kern-.125emX}}
\begin{document}

\IEEEspecialpapernotice{(Invited Paper)}

\title{Integrated Super-Resolution Sensing and Communication with 5G NR Waveform: Signal Processing with Uneven CPs and Experiments}
\author{\IEEEauthorblockN{Chaoyue Zhang\IEEEauthorrefmark{1}, Zhiwen Zhou\IEEEauthorrefmark{1}, Huizhi Wang\IEEEauthorrefmark{1}, and Yong Zeng\IEEEauthorrefmark{1}\IEEEauthorrefmark{2}}
	
\IEEEauthorblockA{\IEEEauthorrefmark{1}National Mobile Communications Research Laboratory, Southeast University, Nanjing 210096, China}
\IEEEauthorblockA{\IEEEauthorrefmark{2}Purple Mountain Laboratories, Nanjing 211111, China}

	\IEEEauthorblockA{Email: 220211013@seu.edu.cn, 220220790@seu.edu.cn, wanghuizhi@seu.edu.cn, yong\underline{~}zeng@seu.edu.cn.}
}
\maketitle
\begin{abstract}
Integrated sensing and communication (ISAC) is a promising technology to simultaneously provide high-performance wireless communication and radar sensing services in future networks. 
In this paper, we propose the concept of \emph{integrated super-resolution sensing and communication} (ISSAC), which uses super-resolution algorithms in ISAC systems to achieve extreme sensing performance for those critical parameters, such as delay, Doppler, and angle of the sensing targets. 
Based on practical fifth generation (5G) New Radio (NR) waveforms, the signal processing techniques of ISSAC are investigated and prototyping experiments are performed to verify the achievable performance.  
To this end, we first study the effect of uneven cyclic prefix (CP) lengths of 5G NR orthogonal frequency division multiplexing (OFDM) waveforms on various sensing algorithms. 
Specifically, the performance of the standard Periodogram based radar processing method, together with the two classical super-resolution algorithms, namely, MUltiple SIgnal Classification (MUSIC) and Estimating Signal Parameter via Rotational Invariance Techniques (ESPRIT) are analyzed in terms of the delay and Doppler estimation. 
To resolve the uneven CP issue, a new structure of steering vector for MUSIC and a new selection of submatrices for ESPRIT are proposed.
Furthermore, an ISSAC experiment platform is setup to validate the theoretical analysis, and the experimental results show that the performance degradation caused by unequal CP length is insignificant and high-resolution delay and Doppler estimation of the target can be achieved with 5G NR waveforms.

\end{abstract}

\section{Introduction}
Integrated sensing and communication (ISAC) is an emerging technology that received significant research interest recently\cite{liu2022integrated}, which aims to provide high-performance wireless communication and ubiquitous radar sensing service simultaneously. 
Extensive research efforts have been devoted to the theoretical analysis and optimization in terms of waveform design \cite{b59}\cite{waveform}, seamless sensing coverage\cite{coverage}, beamforming\cite{hua2023optimal}, beam alignment~\cite{b74}\cite{b76} and information-theoretical limits analysis~\cite{b64,crb,snr}.
On the other hand, practical signal processing techniques for specific sensing tasks such as target detection or parameter estimation have received relatively less attention for ISAC.

In particular, one of the important tasks for ISAC is to estimate the critical parameters of the sensing targets, such as angle of arrival (AoA), propagation delay, and Doppler, etc. 
For ISAC systems based on orthogonal frequency division multiplexing (OFDM) waveforms, the periodogram\cite{sturm2011gemeinsame}\cite{2014ofdmradar} algorithm is a practical method for parameter estimation.
Besides, spectral-based algorithms such as MUltiple SIgnal Classification (MUSIC)\cite{li2019conditioning} and Estimation of Signal Parameters via Rotational Invariance Techniques (ESPRIT)\cite{b46} are classical algorithms to achieve super-resolution, for which super-resolution was initially defined as a resolution exceeding the Rayleigh limit for AoA estimation\cite{gonen1999subspace}. This concept can be further extended to delay and Doppler estimation, which means a resolution exceeding that achieved by the traditional DFT-based algorithms limited by the bandwidth and duration of the radar signal\cite{radarp}.
Specifically, MUSIC algorithm manipulates the signal autocorrelation matrix to yield noise and signal subspaces, and super-resolution is achieved by exploiting the orthogonality between the two subspaces. 
However, in spite of the super-resolution ability, MUSIC algorithm suffers from extremely high computational complexity for spectral search and it is most effective only for relatively high signal-to-noise ratio (SNR). 
On the other hand, super-resolution is achieved by ESPRIT algorithm via exploiting the rotational invariance of the array structure, thus reducing the computational complexity by avoiding spectrum search. 

In this paper, we propose the concept of integrated super-resolution sensing and communication (ISSAC), where a common waveform is used for both communication and sensing, and super-resolution algorithm is used for radar signal processing to achieve extreme sensing performance. 
In particular, we consider the practical fifth generation (5G) New Radio (NR) OFDM waveforms for ISSAC. 
Note that while OFDM waveforms have been extensively investigated for ISAC systems\cite{2014ofdmradar,delay_and_Doppler,zheng2017super,xie2021performance,isac5g}, these works did not consider implementation in practical communication scenario such as 5G NR.
On the other hand, \cite{henninger2022computationally}\cite{pucci2021performance} applied super-resolution algorithms to 5G NR ISAC system, but the signals were modeled as standard OFDM waveforms, ignoring the existence of uneven cyclic prefix (CP) in practical 5G NR waveforms.
To be more specific, it was specified in 3GPP TS 38.211\cite{3gpp38211} that the lengths of CPs in 5G NR OFDM symbols are different in general. 
For super-resolution sensing algorithms, such an uneven CP length may cause extra phase jumps in the symbol (slow-time) domain, which may lead to severe bias and increased variance in Doppler estimation, especially when the target velocity is high.

To deal with the above issue, three methods are investigated in this paper to mitigate the influence of uneven CPs in 5G NR waveforms. 
The first method ignores the extra phase jump caused by long CPs by applying sensing algorithms as if all OFDM symbols had normal CPs. 
The second method takes into account the existence of long CPs by assuming that all OFDM symbol lengths are equal to the average length of OFDM symbols in a frame.
The third method considers the exact structure of OFDM symbols in 5G NR waveforms.
Furthermore, we set up an ISSAC prototype system based on 5G NR waveform, which includes Universal Software Radio Peripheral (USRP), millimeter Wave (mmWave) phased array, high precision slide rail and metal plate.
The experimental results show that with the properly selected method proposed above, only sutble performance degradation will be caused by unequal CP length when estimating delay and Doppler using 5G NR waveforms.

It is worth mentioning that some relevant experiments on ISAC \cite{An_Experimental_Proof}\cite{An_Experimental_Study} have been performed to verify the trade off between communication and sensing performance.
The authors in \cite{First_Demonstration} introduced MIMO radar based on OFDM waveform for high-resolution synthetic aperture radar (SAR) imaging and data transmission. 
ISAC systems were also built to verify the performance improvements based on time-division mode\cite{Joint_Communication},\cite{Spatial_Modulation} and frequency-division mode\cite{Multifunctional_Transceiver}, respectively. 
Different from such existing works, our current work consider both theoretical analysis and prototyping experiments with super-resolution algorithms based on practical 5G NR waveforms with uneven CPs.

\section{System model}\label{sys model}
\begin{figure}[htbp]
	\setlength{\abovecaptionskip}{0.1cm}
	\setlength{\belowcaptionskip}{0.1cm}
	{\includegraphics[width=0.4\textwidth]{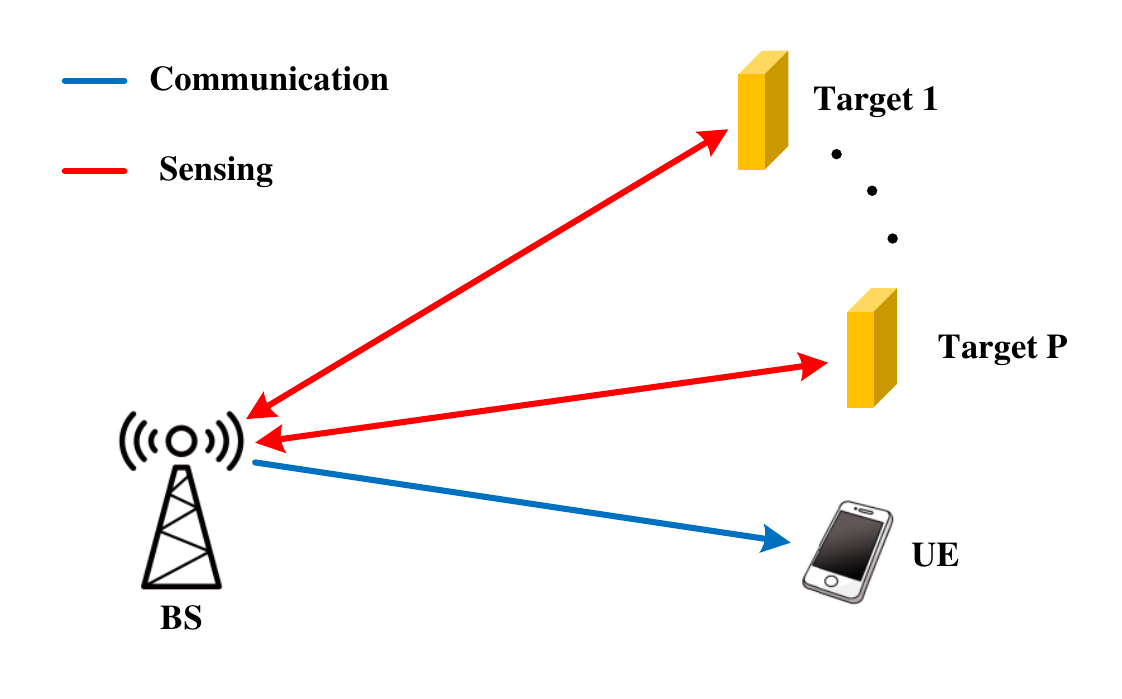}}
	\caption{An ISSAC system based on 5G NR waveform.}
	\label{figplat}
\end{figure}

We consider an ISSAC system with a dual-functional base station (BS), as shown in Fig. \ref{figplat}. 
For simplicity, the BS serves one communication user equipment (UE), and simultaneously senses $P$ targets. The standard 5G NR waveform is considered according to 3GPP TS 38.211\cite{3gpp38211}. 
The corresponding waveform parameters for subsequent prototyping experiments in Section \ref{4 exp and me} are specified in Table \ref{NR Parameters}, where ${f_c}$ is the center frequency, $B$ is the bandwidth, $\mu$ is the subcarrier spacing configuration parameter and $\Delta f$ is the subcarrier spacing. 
Besides, the number of slots per frame and the number of symbols per slot are denoted by $N_{{\rm{slot}}}^{{\rm{frame}}}$ and $N_{{\rm{symb}}}^{{\rm{slot}}}$, respectively. 
Therefore, the number of symbols within each frame is ${N_{{\rm{symb}}}} = N_{{\rm{slot}}}^{{\rm{frame}}}N_{{\rm{symb}}}^{{\rm{slot}}}$.
According to 3GPP TS 38.101\cite{3gpp38201}, the maximum number of Resource Blocks (RB) is denoted as ${N_{{\rm{RB}}}}$, and each RB contains 12 subcarriers. 
Therefore, the total number of subcarriers within each RB is ${N_{{\rm{sc}}}} = 12 \times {N_{{\rm{RB}}}}$. 

\begin{table}[htbp]
	\begin{center}
		\caption{5G NR waveform parameters for ISSAC experiments.}
			\begin{tabular}{|c|c|}
				\hline
			band                                                & FR2 N257                  \\ \hline
				center frequency, ${f_c}$                           & \SI{28}{\giga\hertz}      \\ \hline
				bandwidth, $B$                                      & \SI{400}{\mega\hertz}     \\ \hline
				subcarrier spacing configuration, $\mu$             & \SI{3}{}                  \\ \hline
				subcarrier spacing, $\Delta f$                      & \SI{120}{\kilo\hertz}     \\ \hline
				slots per frame, $N_{{\rm{slot}}}^{{\rm{frame}}}$   & 80                        \\ \hline
				symbols per slot, $N_{{\rm{symb}}}^{{\rm{slot}}}$   & 14                        \\ \hline
				number of symbols per frame, $N_{{\rm{symb}}}$      & 1120                      \\ \hline
				number of RBs, $N_{{\rm{RB}}}$                      & 264                       \\ \hline
				number of subcarriers, $N_{{\rm{sc}}}$              & 3168                      \\ \hline
                frame duration                                      & \SI{10}{\milli\second}    \\ \hline
                slot duration                                       & \SI{125}{\micro\second}   \\ \hline
                long CP length                                      & \SI{9.4}{\micro\second}   \\ \hline
                normal CP length                                    & \SI{8.9}{\micro\second}   \\ \hline
			\end{tabular}
		\label{NR Parameters}
	\end{center}
\end{table}
In 5G NR, the lengths of CP for different OFDM symbols might be different. 
Therefore, we classify symbols into two types, namely symbols with normal CP and symbols with ``long CP". 
Note that the term ``long CP" is different from the concept of ``extended CP" in 5G NR. 
Extended CP is only used in the configuration of $\mu  = 2$, whereas ``long CP" exists in 5G NR frames for all configurations.

\begin{figure}[htbp]
	\setlength{\abovecaptionskip}{0.1cm}
	\setlength{\belowcaptionskip}{0.1cm}
{\includegraphics[width=0.48\textwidth]{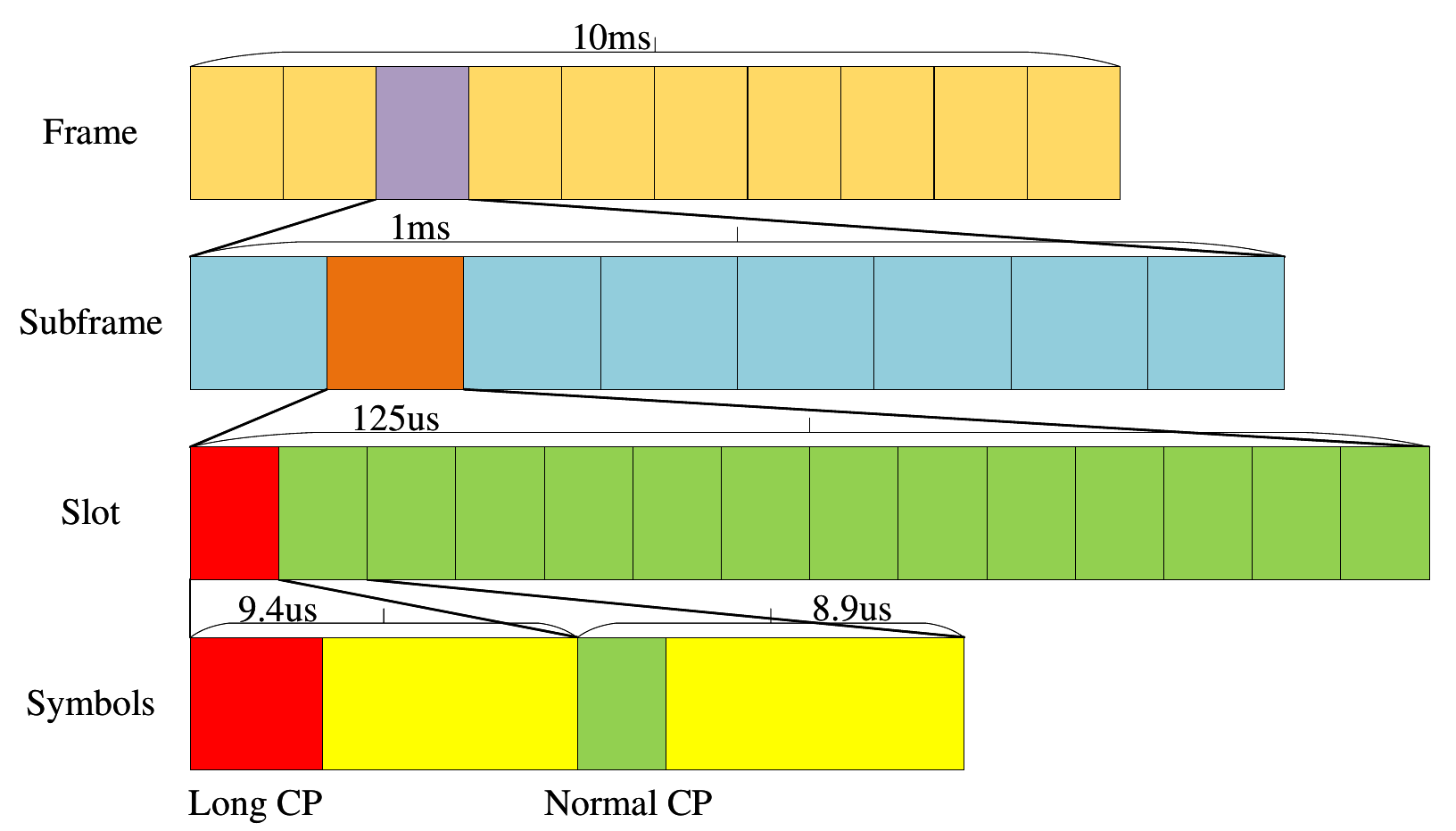}}
	\caption{5G NR frame structure when $\mu  = 3$.}
	\label{fig2}
\end{figure}

The structure of the considered 5G NR frame is illustrated in Fig. \ref{fig2}.

The time-domain waveform of one 5G NR frame can be expressed as
\begin{equation}
\setlength\abovedisplayskip{2pt}
		\setlength\belowdisplayskip{2pt}
	\begin{aligned}
	s\left( t \right) = \sum\limits_{l = 0}^{{N_{{\rm{symb}}}}-1} {{s_l}\left( t \right)}, 
 	\end{aligned}\label{eq1}
\end{equation}
where ${s_l}\left( t \right)$ is the waveform of the $l$th OFDM symbol including CP, which can be expressed as 
\begin{equation}
\setlength\abovedisplayskip{2pt}
		\setlength\belowdisplayskip{2pt}
	\begin{aligned}
	{s_l}\left( t \right) = \left\{ {\begin{array}{*{20}{c}}
			{{{\bar s}_l}\left( t \right)},&{{t_{{\rm{start}},l}} \le t < {t_{{\rm{start}},l}} + {T_{{\rm{symb}},l}}},\\
			0,&{{\rm{otherwise}}},
	\end{array}} \right.
 	\end{aligned}\label{eqslt}
\end{equation}
and
\begin{equation}
\setlength\abovedisplayskip{2pt}
		\setlength\belowdisplayskip{2pt}
	\begin{aligned}
	{\bar s_l}\left( t \right) = \sum\limits_{k = 0}^{{N_{{\rm{sc}}}} - 1} {{b_{k,l}}{e^{j2\pi k\Delta f\left( {t - {N_{{\rm{CP}},l}}{T_c} - {t_{{\rm{start}},l}}} \right)}}},  \label{eqsltbar}
 	\end{aligned}
\end{equation}
where $b_{k,l}$ is the information-bearing symbols for communication at the $k$th subcarrier of the $l$th symbol. Besides, 
\begin{equation}
\setlength\abovedisplayskip{2pt}
		\setlength\belowdisplayskip{2pt}
    {T_{{\rm{symb}},l}} = \left( {{N_u} + {N_{{\rm{CP}},l}}} \right){T_C}
    \label{tsymbl}
\end{equation}
is the length of the $l$th OFDM symbol including CP, with ${{N_u} = 2048\kappa  \cdot {2^{ - \mu }}}$ and ${T_C} = 1/\left( {\Delta f_{\max} \times {N_f}} \right)$ denoting the data length of OFDM symbol and the basic time unit for NR, respectively, with ${N_f = 4096}$ and $\Delta f_{\max}=\SI{480}{\kilo\hertz}$. Thus ${T_C}=\SI{2.0345}{\nano\second}$.
 ${{N_{{\rm{CP}},l}}}$ is the CP length of the $l$th OFDM symbol, which can be expressed as
\begin{equation}
\setlength\abovedisplayskip{2pt}
		\setlength\belowdisplayskip{2pt}
	\begin{aligned}
	{N_{{\rm{CP}},l}} = \left\{ {\begin{array}{*{20}{c}}
			{144\kappa  \cdot {2^{ - \mu }} + 16\kappa }&{}&{l = 0}&{{\rm{or}}}&{l = 7 \cdot {2^\mu }},\\
			{144\kappa  \cdot {2^{ - \mu }}}&{}&{l \ne 0}&{{\rm{and}}}&{l \ne 7 \cdot {2^\mu }},
	\end{array}} \right.
 	\end{aligned}\label{eqncpl}
\end{equation}
where ${\kappa  = T_{\rm{s}}}/{T_{\rm{c}} = 64}$ is a constant, with ${T_{\rm{s}}} = 1/\left( {\Delta {f_{{\rm{ref}}}} \times {N_{{\rm{f}},{\rm{ref}}}}} \right)$ denoting the basic time unit for LTE, in which $\Delta {f_{{\rm{ref}}}} = \SI{15}{\kilo\hertz}$ and ${N_{{\rm{f}},{\rm{ref}}}} = 2048$. 
It is observed from (\ref{eqncpl}) that the CP lengths are uneven in 5G NR. 
Specifically, for the considered configuration of $\mu=3$, there is one OFDM symbol with long CP in every 56 OFDM symbols. Therefore, $\eta  = 7 \times {2^\mu }$ is defined as the separation for adjacent long CPs. 
Besides, ${t_{{\rm{start}},l}}$ in (\ref{eqslt}) denotes the start time of the $l$th OFDM symbol, given by
\begin{equation}
\setlength\abovedisplayskip{2pt}
		\setlength\belowdisplayskip{2pt}
	\begin{aligned}
{t_{{\rm{start}},l}} = \left\{ {\begin{array}{*{20}{c}}
		0&{l = 0},\\
		{{t_{{\rm{start}},l - 1}} + {T_{{\rm{symb}},l - 1}}}&{{\rm{otherwise}}}.
\end{array}} \right.\label{eqtstart}
	\end{aligned}
\end{equation}

With the transmitted signal (\ref{eq1}), the received echo signal by the BS due to target reflection can be expressed as
\begin{equation}
\setlength\abovedisplayskip{2pt}
		\setlength\belowdisplayskip{2pt}
	\begin{aligned}
y\left( t \right) = \sum\limits_{p = 1}^P {{\alpha _p}s\left( {t - {\tau _p}} \right){e^{j2\pi f_pt}}}  + w\left( t \right),
\label{eqyt}
	\end{aligned}
\end{equation}
where $\alpha_p$ is the reflection coefficient that includes the impact of radar cross section (RCS) of the $p$th target, $\tau_p$ and $f_p$ denote the propagation delay and the Doppler frequency of the $p$th target, respectively. 
Once these two parameters are estimated, the target distance and radial velocity can be obtained as $d_p = {c_0}{\tau _p}/2$ and $v_p = {{\left( {f_p{c_0}} \right)} \mathord{\left/
 {\vphantom {{\left( {f_p{c_0}} \right)} {\left( {2{f_C}} \right)}}} \right.
 \kern-\nulldelimiterspace} {\left( {2{f_C}} \right)}}$, where ${c_0}$ denotes the speed of light.
$w\left( t \right)$ is the independent and identically distributed (i.i.d.) Additive White Gaussian Noise (AWGN) with power spectral density $N_0$.
At the UE side, the communication process is carried out based on the standard OFDM receiver, and the detail will not be discussed here as we focus on the sensing performance of the ISSAC system.

We assume that the maximum delay does not exceed the CP length, so that after CP removal, $y\left( t \right)$ can be rearranged into a $K\times L$ matrix $\bf{Y}$ by performing the Fast Fourier Transform (FFT) algorithm\cite{2014ofdmradar}, with $K$ and $L$ denoting the number of subcarriers and symbols, respectively. Therefore, the $\left( {k,l} \right)$th element of $\bf{Y}$ can be expressed as
\begin{equation}
\setlength\abovedisplayskip{2pt}
		\setlength\belowdisplayskip{2pt}
	\begin{aligned}
	{\bf{Y}}\left( {k,l} \right) = {b_{k,l}}\sum\limits_{p = 1}^P {{\alpha _p}{e^{{\rm{ - }}j2\pi k{\tau _p}\Delta f}}{e^{j2\pi f_p{T_{{\rm{sum}},l}}}}}  + w_{k,l},
 	\end{aligned}\label{eqykl}
\end{equation}
in which ${T_{{\rm{sum}},l}}$ is the cumulative durations at the $l$th symbol, Based on (\ref{tsymbl}) and (\ref{eqncpl}), ${T_{{\rm{sum}},l}}$ can be expressed as
\begin{equation}
\setlength\abovedisplayskip{2pt}
\setlength\belowdisplayskip{2pt}
    \begin{aligned}
                {T_{{\rm{sum}},l}} &= \sum\limits_{i = 0}^l {{T_{{\rm{symb}},i}}}   \\
                &= {\left( {{N_u} + 144\kappa  \times {2^{ - \mu }}} \right){T_C}}l + 16\kappa {T_C}\left\lceil {\frac{l}{\eta }} \right\rceil ,
    \label{tsum}
    \end{aligned}
\end{equation}
where $\left\lceil x \right\rceil\ $denotes the minimum integer no less than $x$.
Furthermore, by dividing the random communication symbols $b_{k,l}$ and substituting (\ref{tsum}) into (\ref{eqykl}), the data matrix ${\bf{\tilde Y}}$ for radar processing can be obtained as:
\begin{equation}
\small
\setlength\abovedisplayskip{2pt}
		\setlength\belowdisplayskip{2pt}
	\begin{aligned}
&{\bf{\tilde Y}}\left( {k,l} \right) = \frac{{\bf{Y}}\left( {k,l} \right)}{{{b_{k,l}}}} = \sum\limits_{p = 1}^P {{\alpha _p}{e^{ - j2\pi k{\tau _p}\Delta f}}{e^{j2\pi f_p{T_{{\rm{sum}},l}}}}}  + \tilde w_{k,l}\\
& = \sum\limits_{p = 1}^P {{\alpha _p}{e^{ - j2\pi k{\tau _p}\Delta f}}{e^{j2\pi f_p{\left( {{N_u} + 144\kappa  \times {2^{ - \mu }}} \right){T_C}}l}}} {e^{j2\pi f_p16\kappa {T_C}\left\lceil {\frac{l}{\eta }} \right\rceil }} \\
& + {{\tilde w}_{k,l}}.
	\end{aligned}\label{eqYkl}
\end{equation}

Different from the standard OFDM model\cite{2014ofdmradar}\cite{b73}, due to the existence of long CP of 5G NR waveforms, there is an additional phase shift ${\Delta \phi \,\,=2\pi f_{p}16\kappa T_C\lceil \frac{l}{\eta} \rceil}$ in (\ref{eqYkl}), which is generally small and no longer increase linearly with the symbol index $l$.
For the waveform specified in Table \ref{NR Parameters}, by substituting $l \le N_{{\rm{symb}}}-1$ and $\Delta f=\SI{120}{\kilo\hertz}$, we have ${\left| \Delta \phi \right|\leqslant {{\pi} /{4}}}$ if ${\left| f_{p} \right|\leqslant \frac{1}{10}\Delta f}$. 
Such an additional phase shift may cause bias in Doppler estimation, which has not been considered in traditional OFDM radar processing. 
In this paper, three methods are investigated to deal with the above issue:
\begin{enumerate}
    \item\label{case1} 
    In the first method, the existence of long CPs in the NR OFDM frame and the extra phase jumps caused by it are not considered at all, which means the lengths of all OFDM symbols are considered equal to the length of OFDM symbols with normal CPs. 
    In this way, ${T_{{\rm{sum}},l}}$ can be expressed as:
    \begin{equation}
        {T_{{\rm{sum}},l}^{\rm{I}}} = \left( {{N_u}+144\kappa  \times {2^{ - \mu }}} \right){T_C}l.\label{tsum1}
    \end{equation}
    \item\label{case2} 
    In the second method, the extra phase introduced by long CPs is approximated by assuming that all symbols have the same length, which is equal to the average symbol length of each OFDM symbol in a frame.
    ${T_{{\rm{sum}},l}}$ is chosen to be the average of all symbol lengths of an NR OFDM frame with duration ${T_{{\rm{frame}}}}=\SI{10}{\milli\second}$:
    \begin{equation}
        {T_{{\rm{sum}},l}^{\rm{II}}} = {T_{{\rm{frame}}}}l/{N_{{\rm{symb}}}}.\label{tsum2}
    \end{equation}
    \item\label{case3} In the third method, the exact structure of the NR OFDM frame with long CPs is considered, as shown in (\ref{tsum}). Considering the different principles for different sensing algorithms, the specific countermeasures taken will vary depending on the chosen algorithm, which will be discussed in detail in the next section.
\end{enumerate}


In the following, three classical sensing algorithms are modified based on the three methods above to mitigate the influence of uneven CPs. 

\section{Super-Resolution sensing algorithm with uneven CP}\label{3principle and simu}
In this section, three classic sensing algorithms, namely Periodogram, MUSIC and ESPRIT are studied to estimate the delay and Doppler parameters based on 5G NR waveforms. 
The Periodogram algorithm serves as the baseline for the super-resolution algorithms in ISSAC system.
Firstly, we introduce the principles and implementation methods of the different algorithms. 
Furthermore, we perform three algorithms on $\bf{\tilde Y}$ to estimate the delay and Doppler parameter, and analyze the impact of uneven CP in 5G NR waveforms on different algorithms, respectively.


\subsection{Periodogram}\label{3A}
In this subsection, the one-dimensional Periodogram\cite{2014ofdmradar} is performed in the delay and Doppler domain as a baseline of super-resolution sensing algorithms.
The FFT and IFFT are applied to each row and column of $\bf{\tilde Y}$, and their Periodogram can be expressed as: 
\begin{equation}
\setlength\abovedisplayskip{2pt}
		\setlength\belowdisplayskip{2pt}
	\begin{aligned}
	{{\bf{\bar y}}_f} &= \sum\limits_{k = 1}^{{N_{{\rm{sc}}}}} {\frac{1}{{{N_{{\rm{sc}}}}{N_{{\rm{symb}}}}}}{{\left| {{\rm{FF}}{{\rm{T}}_{N_{{\rm{symb}}}^{{\rm{FFT}}}}}\left( {{{{\bf{\tilde y}}}_k}} \right)} \right|}^2}}, \\
 {{\bf{\bar y}}_\tau } &= \sum\limits_{l = 1}^{{N_{{\rm{symb}}}}} {\frac{1}{{{N_{{\rm{sc}}}}{N_{{\rm{symb}}}}}}{{\left| {{\rm{IFF}}{{\rm{T}}_{N_{{\rm{sc}}}^{{\rm{FFT}}}}}\left( {{\bf{\tilde y}}}_l^{\rm{T}} \right)} \right|}^2}}, 
 	\end{aligned}
\end{equation}
where ${{\bf{\tilde y}}}_n$ and ${\bf{\tilde y}}_n^{\rm{T}}$ denotes the $n$th row and column of matrix $\bf{\tilde Y}$, respectively. The peaks of ${{\bf{\bar y}}_f}$ and ${{\bf{\bar y}}_\tau }$ correspond to the Doppler and delay of the targets, respectively. $N_{{\rm{symb}}}^{{\rm{FFT}}}$ and $N_{{\rm{sc}}}^{{\rm{FFT}}}$ are the number of points for fast Fourier transforms. 
It is assumed that each symbol in OFDM frame has the same length $T_{\rm{symb}}$, and the Doppler and delay of the target can be solved by the peak index ${n_f}$ and ${n_\tau }$ of the periodogram:
\begin{equation}
\setlength\abovedisplayskip{2pt}
		\setlength\belowdisplayskip{2pt}
\begin{array}{*{20}{c}}
{{f} = \frac{{{n_f}}}{{N_{{\rm{symb}}}^{{\rm{FFT}}}{T_{{\rm{symb}}}}}}},&{\tau  = \frac{{{n_\tau }}}{{N_{{\rm{sc}}}^{{\rm{FFT}}}\Delta f}}}.
\end{array}\label{fd}
\end{equation}

The delay and Doppler resolutions of the Periodogram algorithm are limited by the signal duration and bandwidth. Therefore, a trade-off between resolution and maximum measurement range needs to be considered \cite{2018VNC}. 
Note that in Periodogram, the long CP introduces a slight phase shift in the symbol (slow-time) domain.
Therefore, we design three signal processing methods for Periodogram according to Section \ref{sys model}.
In particular, for method 3, the symbol structure is no longer uniform, so the traditional FFT is no longer applicable. 
We construct matched filter by substituting (\ref{fd}):  
\begin{equation}
\setlength\abovedisplayskip{2pt}
\setlength\belowdisplayskip{2pt}
    {\bf{F}}\left( {n,m} \right) = {e^{\frac{{ - j2\pi n{T_{{\rm{sum,m}}}}}}{{N_{{\rm{symb}}}^{{\rm{FFT}}}T_{{\rm{sum,1}}}^{\rm{I}}}}}},
\end{equation}
where $n,m=1,...,{N_{{\rm{symb}}}^{{\rm{FFT}}}}$.
The periodogram of the matched filter is
 \begin{equation}
     {{\bf{\bar y}}_f} = \sum\limits_{k = 1}^{{N_{{\rm{sc}}}}} {\frac{1}{{{N_{{\rm{sc}}}}{N_{{\rm{symb}}}}}}{{\left| {{\bf{F}} {{{{\bf{\tilde y}}}_k}}} \right|}^2}}.
 \end{equation}

\begin{figure}[htbp]
\centering
	{\includegraphics[width=0.5\textwidth]{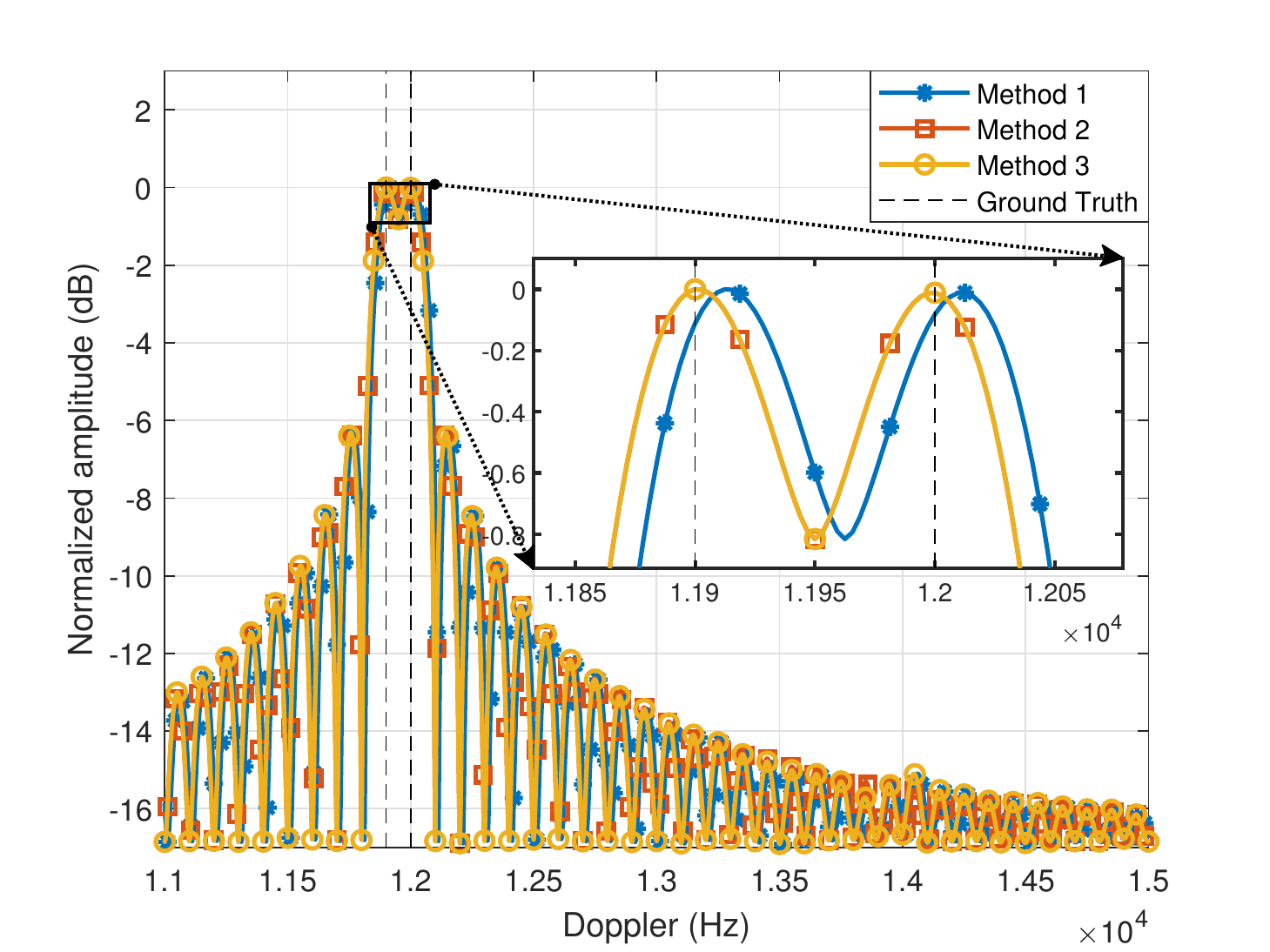}}
	\caption{Periodogram simulation result.}
	\label{perlong}
\end{figure}

Fig. \ref{perlong} shows the simulation results of the three methods of the Periodogram, and the black dotted line represents the ground truth.
There are two targets with the speed of ${v_1} = \SI[per-mode=symbol]{63.75}{\meter\per\second}$ and ${v_2} = \SI[per-mode=symbol]{64.29}{\meter\per\second}$, corresponding to Doppler frequencies $f_1=\SI{11.90}{\kilo\hertz}$ and $f_2=\SI{12.00}{\kilo\hertz}$, respectively.
It can be seen that method 1 leads to obvious bias in estimation, while methods 2 and 3 are unbiased. 
However, due to the poor resolution of the Periodogram, methods 2 and 3 fail to show notable difference between each other.

\subsection{MUSIC}\label{3B}
In this subsection, we study the classic MUSIC algorithm on the ${{N_{{\rm{sc}}}}\times {N_{{\rm{symb}}}}}$ data matrix ${\bf{\tilde Y}}$ and analyze the effect of uneven CPs on the estimation results.
Due to the limited length of measurement data and high correlation between reflected signals, the covariance matrix of ${\bf{\tilde Y}}$ may be rank deficient.
Therefore, we perform modified spatial smoothing preprocessing (MSSP) \cite{Superresolution_techniques_for_time_domain} in the subcarrier domain. Note that no preprocessing in the symbol domain is performed to avoid the destruction of the uneven structure between symbols. To be more specific,
${N_{{\rm{sub}}}}$ submatrices of size $L \times {N_{{\rm{symb}}}}$ are selected in all the ${N_{{\rm{sc}}}}$ subcarriers, where ${N_{{\rm{sub}}}}{\rm{ = }}{N_{{\rm{sc}}}} - L + 1$, $L = \rho {N_{{\rm{sc}}}}$ and $\rho=0.4 $ is the smoothing constant, and the $L \times {N_{{\rm{sub}}}}{N_{{\rm{symb}}}}$ data matrix after MSSP can be expressed as:
\begin{equation} 		
{{{\bf{\hat Y}}}}{\rm{ = }}\left( {\begin{array}{*{20}{c}}
{{{{\bf{\tilde y}}}_1}}& \cdots &{{{{\bf{\tilde y}}}_{{N_{{\rm{sub}}}}}}}\\
 \vdots & \ddots & \vdots \\
{{{{\bf{\tilde y}}}_L}}& \cdots &{{{{\bf{\tilde y}}}_{L{\rm{ + }}{N_{{\rm{sub}}}}}}}
\end{array}} \right).
\label{mssp}
\end{equation}

Furthermore, the eigenvalue decomposition is performed on the covariance matrix of Doppler and delay ${{\bf{R}}_f}$ and ${{\bf{R}}_\tau }$, respectively:
\begin{equation}
\setlength\abovedisplayskip{2pt}
\setlength\belowdisplayskip{2pt}
\begin{aligned}
    {{\bf{R}}_f} &= \frac{{{{{\bf{\tilde Y}}}^{\rm{H}}}{\bf{\tilde Y}}}}{{{N_{{\rm{sc}}}}}} = {\bf{E}}_s^f{\bf{\Lambda }}_s^f{\left( {{\bf{E}}_s^f} \right)^{\rm{H}}} + {\bf{E}}_n^f{\bf{\Lambda }}_n^f{\left( {{\bf{E}}_n^f} \right)^{\rm{H}}},\\
    {{\bf{R}}_\tau }{\rm{ }} &= \frac{{{\bf{\hat Y}}{{{\bf{\hat Y}}}^{\rm{H}}}}}{{{N_{{\rm{sub}}}}{N_{{\rm{symb}}}}}} = {\bf{E}}_s^\tau {\bf{\Lambda }}_s^\tau {\left( {{\bf{E}}_s^\tau } \right)^{\rm{H}}} + {\bf{E}}_n^\tau {\bf{\Lambda }}_n^\tau {\left( {{\bf{E}}_n^\tau } \right)^{\rm{H}}},
\end{aligned}
\label{eqrxxdop}
\end{equation}
where ${{\bf{\Lambda}}_s^q}$ denotes the diagonal matrix with respect to the $P$ largest eigenvalues, and ${{\bf{E}}_s^q}$ and ${{\bf{E}}_n^q}$ denote the signal subspace and noise subspace respectively, with $q\in\{\tau,f\}$. 

Therefore, the MUSIC spectrum can be obtained as:
\begin{equation}
\setlength\abovedisplayskip{2pt}
		\setlength\belowdisplayskip{2pt}
\begin{aligned}
P_f^{{\rm{MUSIC}}}\left( f \right) &= \frac{1}{{{{\left( {{{\bf{x}}^f}} \right)}^{\rm{H}}}{\bf{E}}_n^f{{\left( {{\bf{E}}_n^f} \right)}^{\rm{H}}}{{\bf{x}}^f}}},
\\
P_\tau ^{{\rm{MUSIC}}}\left( \tau  \right) &= \frac{1}{{{{\left( {{{\bf{x}}^\tau }} \right)}^{\rm{H}}}{\bf{E}}_n^\tau {{\left( {{\bf{E}}_n^\tau } \right)}^{\rm{H}}}{{\bf{x}}^\tau }}},
\end{aligned}
\end{equation}
\begin{equation}
\begin{aligned}
{{\bf{x}}^f} &= \left[ {x_0^f,...,x_{{N_{{\rm{symb}}}} - 1}^f} \right]^{\rm{T}},
\\
{{\bf{x}}^\tau } &= \left[ {x_{0,}^\tau ,...,x_{L - 1}^\tau } \right]^{\rm{T}},
\end{aligned}
\end{equation}
where  $x_{l}^f = {e^{j2\pi {f}{T_{{\rm{symb}},l}}}}$ and $x_{k}^\tau  = {e^{ - j2\pi k\tau\Delta f }}$ denote the $l$th and $k$th element of the Doppler and delay steering vectors, respectively. 
In the following, the effect of uneven CP on MUSIC algorithm is analyzed in the Doppler domain. 
Three Doppler steering vector are designed corresponding to the three methods in Section \ref{sys model}, where we have ${T_{{\rm{symb}},l}} = {T_{{\rm{sum}},l}}$ for the third method.  

\begin{figure}[htbp]
	\setlength{\abovecaptionskip}{0.1cm}
	\setlength{\belowcaptionskip}{0.1cm}
	{\includegraphics[width=0.5\textwidth]{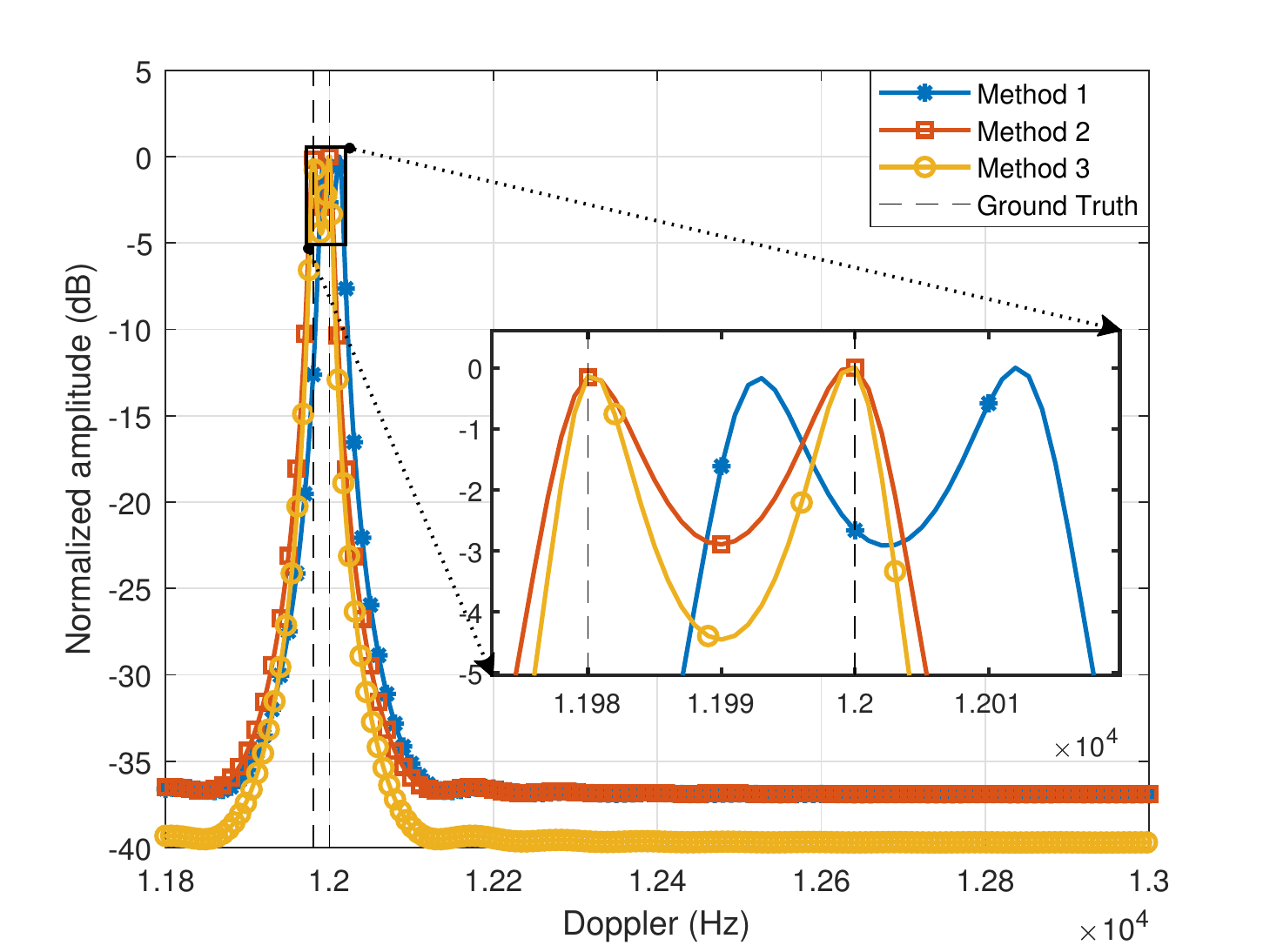}}
	\caption{MUSIC simulation result.}
	\label{figmusic}
\end{figure}

Fig. \ref{figmusic} shows the simulation result of MUSIC algorithm based on three different methods above, where the black dotted line is the true Doppler frequency. 
There are two targets with the speed of ${v_1} = \SI[per-mode=symbol]{64.18}{\meter\per\second}$ and ${v_2} = \SI[per-mode=symbol]{64.29}{\meter\per\second}$, corresponding to Doppler frequencies $f_1=\SI{11.98}{\kilo\hertz}$ and $f_2=\SI{12.00}{\kilo\hertz}$, respectively.
As can be seen from the figure, if the structure of long CP is not considered, severe deviations occur at the MUSIC spectrum peaks when the Doppler frequencies of the two targets are close. 
Method 2 can reduce the deviation of Doppler estimation, but the resolution of the two spectrum peaks is still poor. 
Furthermore, with method 3, the orthogonality between the steering vectors and the noise subspace is enhanced, so the estimation result is accurate with the highest resolution.
\subsection{ESPRIT}\label{3C}
In conventional OFDM radar, the ESPRIT algorithm is used to estimate delay based on the rotational invariance between OFDM signal subcarriers. 
It can also be used to estimate the Doppler by exploiting the uniform structure between multiple OFDM symbols\cite{delay_and_Doppler}. 

The ESPRIT algorithm first needs to select two sub-sequences with a fixed phase shift from the signal subspace, that is, sub-sequences with rotational invariance. This phase shift is called the rotational invariant factor, and this factor can be used to estimate the frequency shift. 
Since no spectrum peak search is required, its complexity is much lower than that of MUSIC algorithm. 
However, ESPRIT algorithm requires strictly uniform structure between elements of sub-sequences.

Similar to MUSIC algorithm, the Doppler and delay domain covariance matrix are obtained according to (\ref{eqrxxdop}). 
The eigenvalue decomposition is performed to obtain the signal subspace ${\bf{E}}_s^f $ and ${\bf{E}}_s^\tau $. 
After that, we select two $\left( N_{\rm symb}-1 \right) \times P$ submatrices from the Doppler signal subspace ${\bf{E}}_s^f $ and two $\left(N_{\rm sc}-1\right) \times P$ submatrices from the delay signal subspace ${\bf{E}}_s^\tau $ to maintain rotational invariance.
\begin{equation}\label{eqjtt}
\setlength\abovedisplayskip{2pt}
\setlength\belowdisplayskip{2pt}
\begin{aligned}
    {{\bf{J}}_{f,1}} &= {\left[ {{\bf{E}}_{{\rm{s,1}}}^f,...,{\bf{E}}_{{\rm{s,}}{N_{{\rm{symb}}}} - 1}^f} \right]^{\rm{T}}},
{{\bf{J}}_{f,2}} = {\left[ {{\bf{E}}_{{\rm{s,2}}}^f,...,{\bf{E}}_{{\rm{s,}}{N_{{\rm{symb}}}}}^f} \right]^{\rm{T}}}\\
{{\bf{J}}_{\tau ,1}} &= {\left[ {{\bf{E}}_{{\rm{s,1}}}^\tau ,...,{\bf{E}}_{{\rm{s,}}{N_{{\rm{sc}}}} - 1}^\tau } \right]^{\rm{T}}},
{{\bf{J}}_{\tau ,2}} = {\left[ {{\bf{E}}_{{\rm{s,2}}}^\tau ,...,{\bf{E}}_{{\rm{s,}}{N_{{\rm{sc}}}}}^\tau } \right]^{\rm{T}}},
\end{aligned}
\end{equation}
where ${\bf{E}}_{{\rm{s}},i}^q$ denotes the $i$th row of the signal subspace. 
Based on the rotational invariance of uniform structure, the rotation invariant matrices of two dimensions are obtained from the submatrices as ${\bf{\Phi }}{\rm{ = }}{\left( {{\bf{J}}_{f,2}^{\rm{H}}{{\bf{J}}_{f,2}}} \right)^{ - 1}}{\bf{J}}_{f,2}^{\rm{H}}{{\bf{J}}_{f,1}}$ and ${\bf{\Psi }}{\rm{ = }}{\left( {{\bf{J}}_{\tau ,2}^{\rm{H}}{{\bf{J}}_{\tau ,2}}} \right)^{ - 1}}{\bf{J}}_{\tau ,2}^{\rm{H}}{{\bf{J}}_{\tau ,1}}$.
Then, the delay and Doppler of multiple targets can be solved by the eigenvalue of ${\bf{\Phi }}{\rm{ = }}{\bf{U}}_f^{ - 1}{{\bf{D}}_f}{{\bf{U}}_f}$ and ${\bf{\Psi }}{\rm{ = }}{\bf{U}}_\tau ^{ - 1}{{\bf{D}}_\tau }{{\bf{U}}_\tau }$, where 
\begin{equation}
\setlength\abovedisplayskip{2pt}
\setlength\belowdisplayskip{2pt}
\begin{aligned}
    {{\bf{D}}_f}&{\rm{ = diag}}\left( {{e^{ - j2\pi f_1{T_{{\rm{symb}}}}}}, \cdots ,{e^{ - j2\pi f_P{T_{{\rm{symb}}}}}}} \right),\\
    {{\bf{D}}_\tau }&{\rm{ = diag}}\left( {{e^{ - j2\pi {\tau _1}\Delta f}}, \cdots ,{e^{ - j2\pi {\tau _P}\Delta f}}} \right).
\end{aligned}
\end{equation}

However, the uneven CP of NR waveform may destroy the uniform structure between symbols, and thus degrade the accuracy of the Doppler estimation using ESPRIT algorithm. 
Similar to the analysis in Section \ref{sys model}, three different method are considered for ${{\bf{D}}_f}$. 
In method 3, we propose a non-uniform ESPRIT solution, which is able to eliminate the effect of uneven CP while maintaining the rotation invariance of the two selected subarrays by removing the elements associated with the long CP symbols from the signal subspace.
To be more specific, ${{\bf{J}}_{f,1}}$ is selected by removing the $\left({\eta n{\rm{ + }}\eta }\right)$th row from ${\bf{E}}_{\rm{s}}^f$, and ${{\bf{J}}_{f,2}}$ is obtained by removing the $\left({\eta n{\rm{ + }}1 }\right)$th row from ${\bf{E}}_{\rm{s}}^f$, where $n=\left( 0,...,N_{\rm symb}/\eta -1 \right)$.

\begin{figure}[htbp]
\setlength\abovedisplayskip{2pt}
\setlength\belowdisplayskip{2pt}
	\centering
	\subfloat[Conventional ESPRIT]{\includegraphics[width=1\columnwidth]{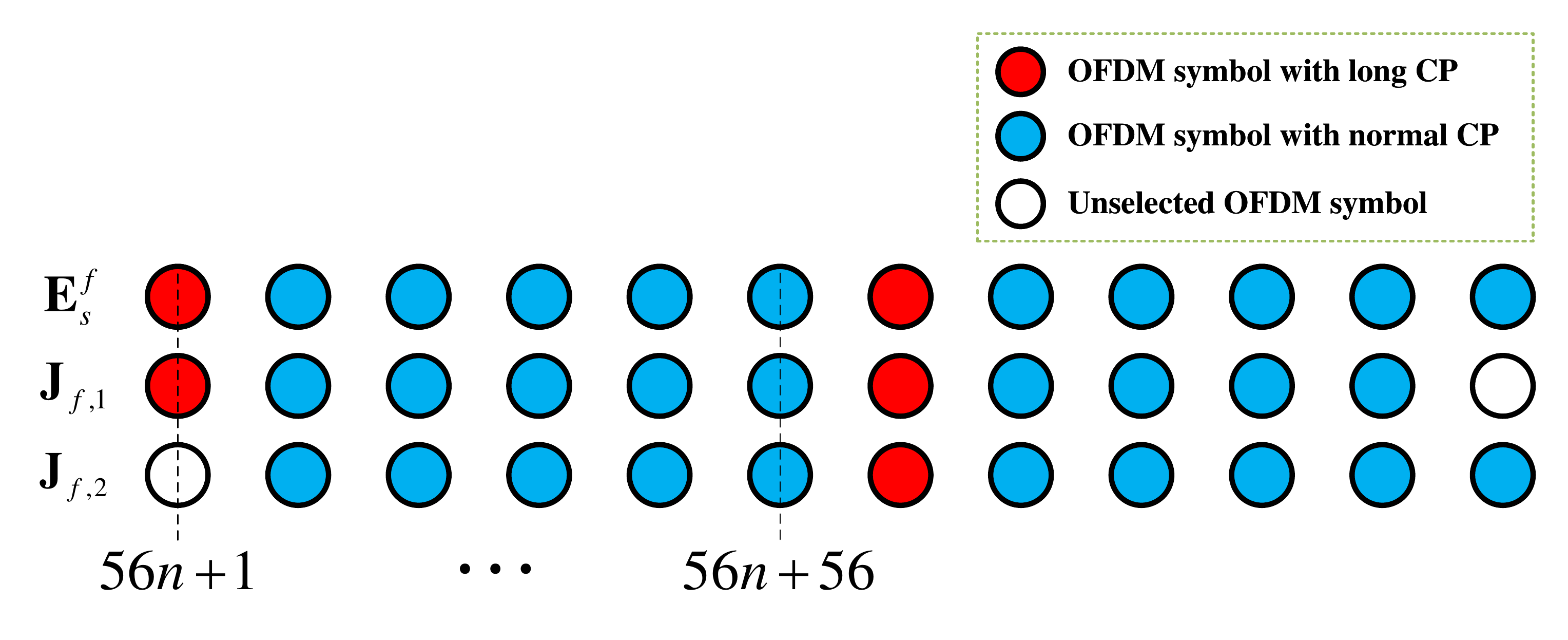}}\hspace{5pt}\\
	\subfloat[Proposed non-uniform ESPRIT]{\includegraphics[width=1\columnwidth]{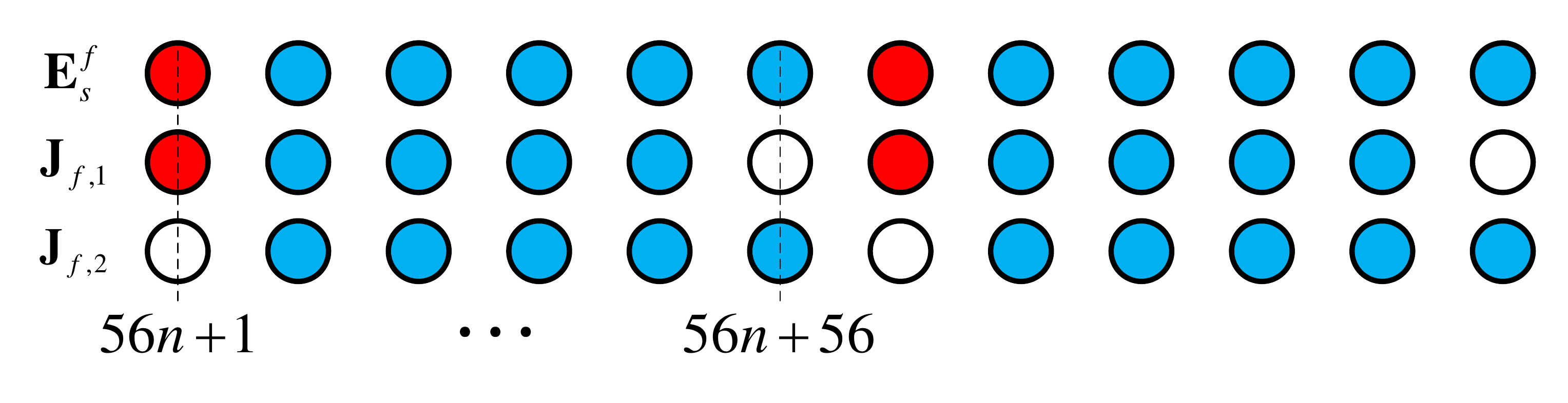}}\hspace{5pt}

	\caption{The selected signal subspace submatrix of conventional ESPRIT versus the proposed non-uniform ESPRIT.} \label{figNon-uniformESPRIT}
\end{figure}

Fig. \ref{figNon-uniformESPRIT} compares the conventional and proposed signal subspace submatrix of ESPRIT algorithm, denoted by the upper part and the lower part, respectively. Each row represents the transposed matrix of ${\bf{E}}_s^f$, and each circle represents a row of ${\bf{E}}_s^f$. The red circle is the row where the long CP is located in the signal subspace, the blue circle is the row of other OFDM symbols and the white circle is the removed row of ${\bf{E}}_s^f$.

\begin{figure}[h]
\centering
	{\includegraphics[width=0.49\textwidth]{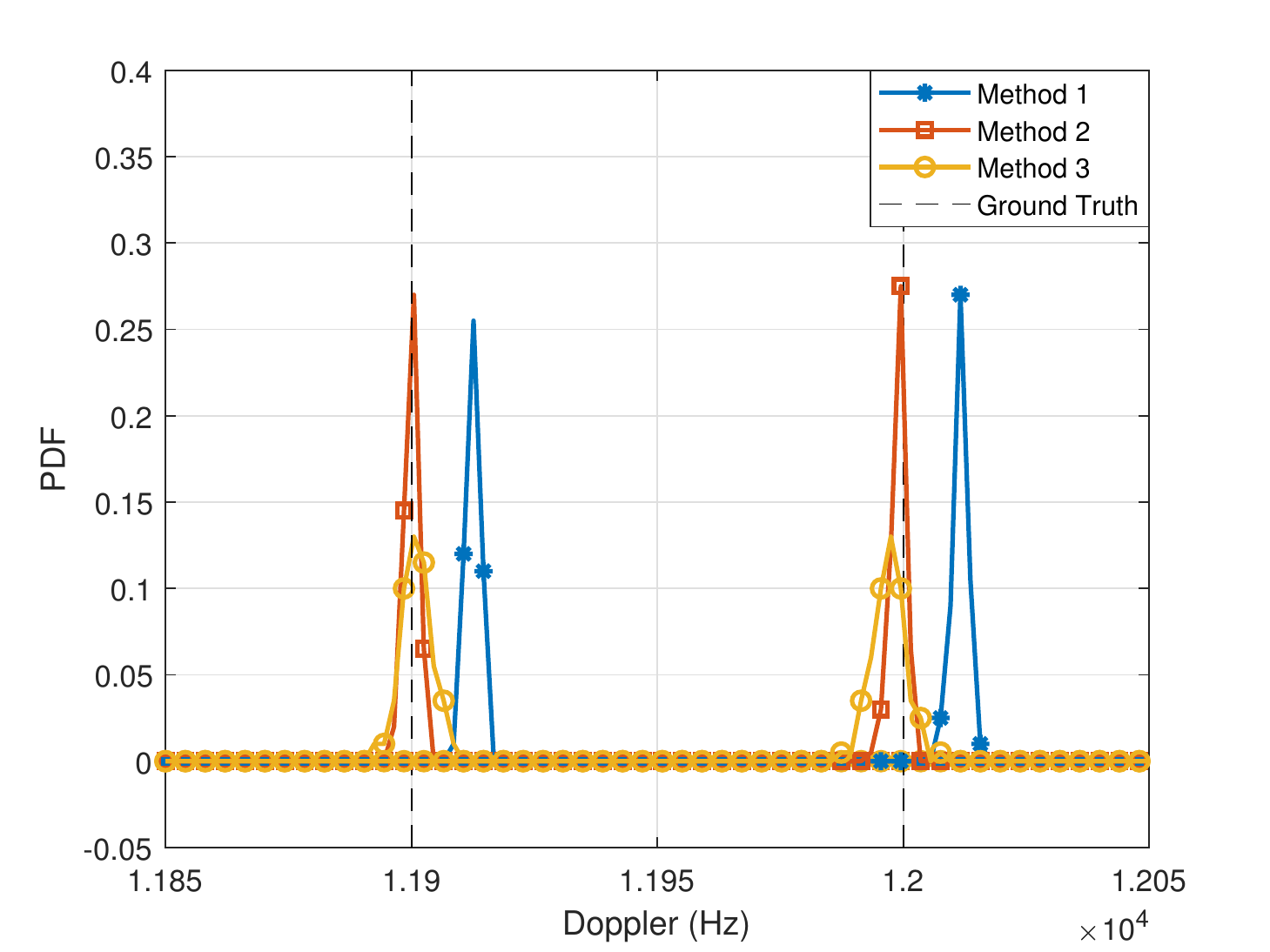}}
\caption{ESPRIT simulation result.}
	\label{figESPRIT}
\end{figure}

Fig. \ref{figESPRIT} shows the estimation result of ESPRIT algorithm running 100 times in each method at SNR of \SI{15}{\dB}, where the black dashed line is the true Doppler frequency. 
There are two targets with the speed of ${v_1} = \SI[per-mode=symbol]{63.75}{\meter\per\second}$ and ${v_2} = \SI[per-mode=symbol]{64.29}{\meter\per\second}$, corresponding to Doppler frequencies $f_1=\SI{11.90}{\kilo\hertz}$ and $f_2=\SI{12.00}{\kilo\hertz}$, respectively.
It can be seen that there is a significant deviation between the estimation result and the true value for method 1, when long CP is not considered. 
Besides, using average symbol length can improve the estimation accuracy. 
Furthermore, considering the effect of long CP in method 3, the accuracy can be increased as well, but the variance is larger than method 2 due to the reduction of array size and low SNR.
This shows that in ESPRIT algorithm, the loss of array size has a greater impact on the algorithm than the loss of rotation invariance.
Therefore, method 2 is used for subsequent experimental data processing.
We do not consider the spatial smoothing before Doppler estimation because of the existence of long CP, which is similar to method 3 considering non-uniform smoothing matrices.

\section{Prototyping experiments and  results}\label{4 exp and me}
\subsection{Experimental Setup}
\begin{figure}[tbp]
\centering
	{\includegraphics[width=0.4\textwidth]{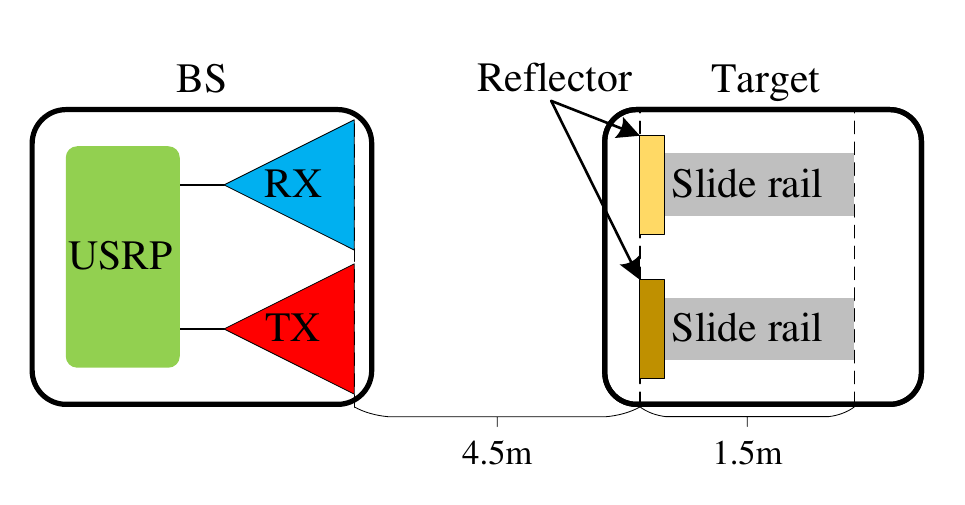}}
	\caption{Experimental setup.}
	\label{figexpset}
\end{figure}

In this section, we validate our developed results with
prototyping experiments. 
The ISSAC experimental platform consists of a BS and a pair of high-precision slide rail with fixed metal plates as the sensing targets, as shown in Fig. \ref{figexpset}.

\begin{figure}[htbp]
	\setlength{\abovecaptionskip}{0.1cm}
	\setlength{\belowcaptionskip}{0.1cm}
\centering
	{\includegraphics[width=0.4\textwidth]{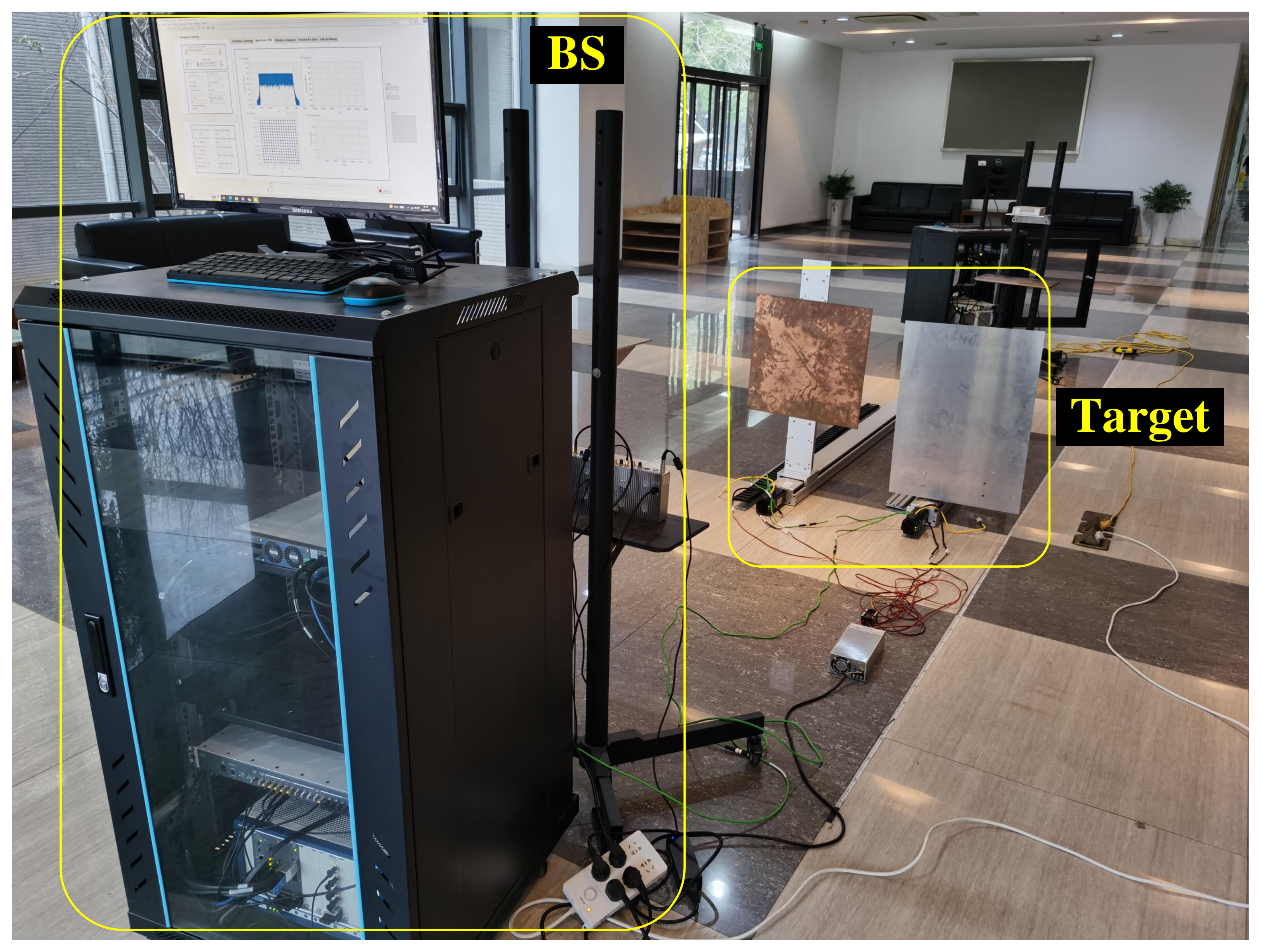}}
	\caption{Prototyping experiment scenarios for ISSAC.}
	\label{figexpshoot}
\end{figure}

A PXIe 8881 is used as a controller at the BS side, which is responsible for 5G NR waveform configuration and generation, radar signal processing, etc. 
Universal Software Radio Peripheral (USRP) X410 is used for the transmission and reception of the Intermediate Frequency (IF) signals. 
The working frequency of X410 ranges from \SI{1}{\mega\hertz} to \SI{7.2}{\giga\hertz}, and the maximum supported bandwidth is \SI{400}{\mega\hertz}. An IF frequency of \SI{2.8}{\giga\hertz} and a bandwidth of \SI{400}{\mega\hertz} are chosen.
Four RF channels are available on the X410 and two of them are connected to a pair of $8 \times 8$ millimeter wave (mmWave) phased arrays with embedded mmWave up and down converters. The up and down converters convert the \SI{2.8}{\giga\hertz} IF signals to and from \SI{28}{\giga\hertz} mmWave signals. For this work, we focus on range and Doppler sensing, and the mmWave beams are configured to point towards the targets. The equivalent isotropically radiated power (EIRP) is configured to be \SI{44}{dBm}.
As shown in Fig. \ref{figexpset} and Fig. \ref{figexpshoot}, two \SI{1.5}{\meter}-long slide rails are placed \SI{4.5}{\meter} away from the BS. Two metal plates are placed on the slide rails with varying positions to mimic the sensing targets.
In the following, the delay and Doppler of the targets are estimated. 

\subsection{Delay Estimation}
We first control the sliding rails to place the targets at a set of fixed distances shown in Table \ref{Distance}.
The BS sends a downlink NR frame to initiate the measurement. 
After finishing the measurement, the BS implements the three sensing algorithms discussed in Section \ref{3principle and simu} to process the received data. 

\begin{figure*}[b]
	\centering
	\subfloat[]{\label{figPER_TAO}
        \includegraphics[width=0.32\textwidth]{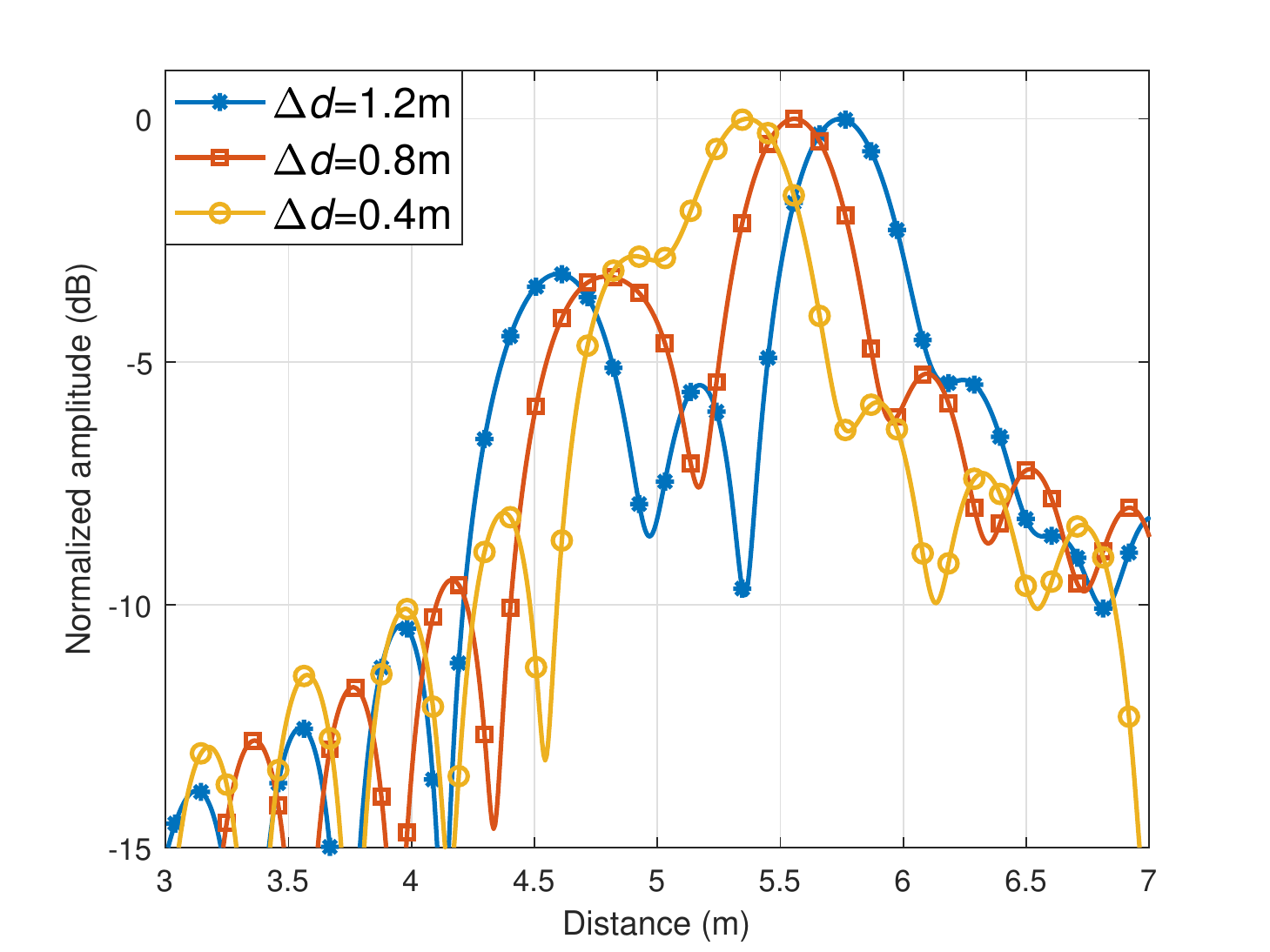}
		}
	\hfil
	\subfloat[]{\includegraphics[width=0.32\textwidth]{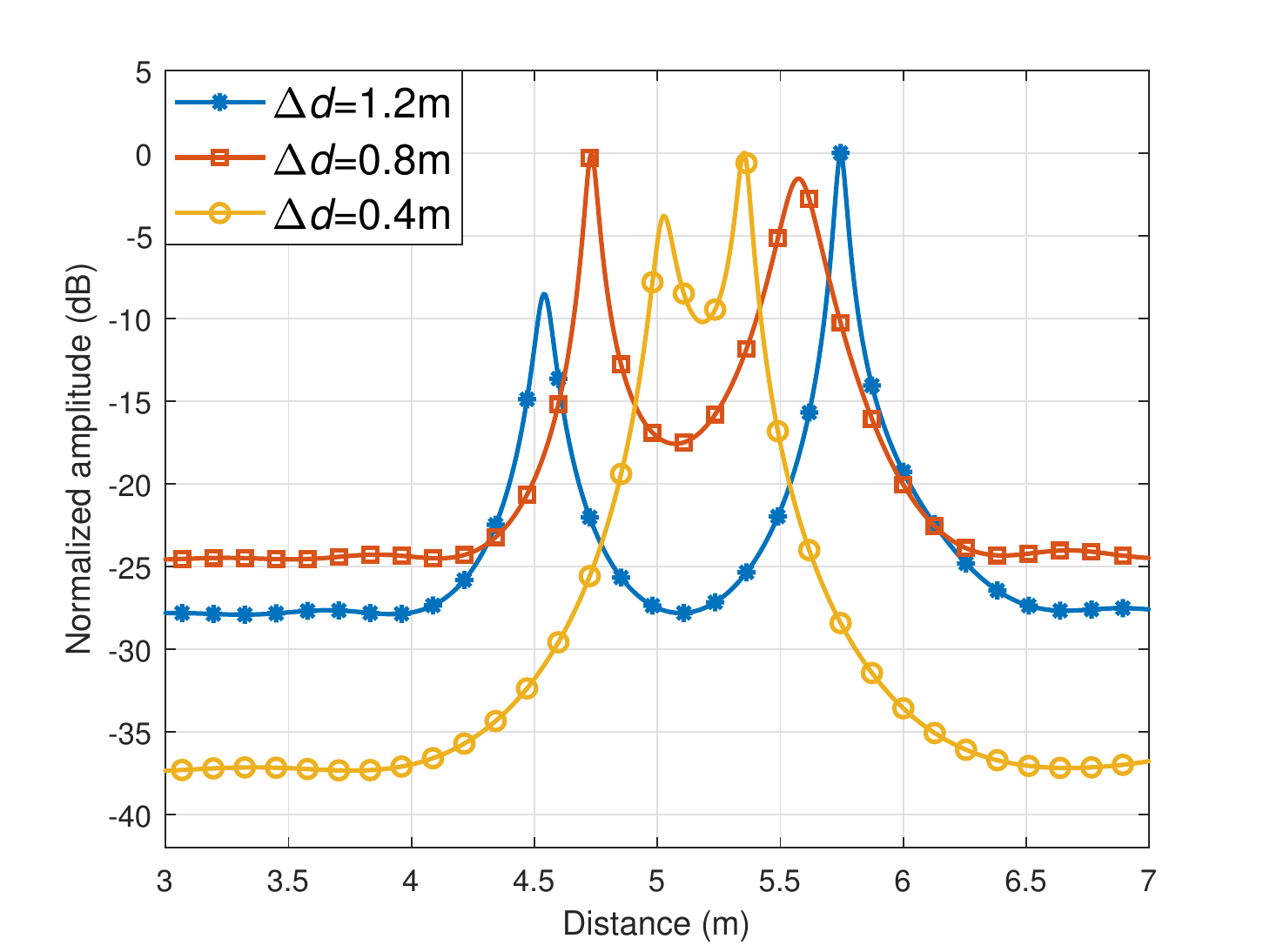}
		\label{figMUSIC_TAO}}
        \hfil
	\subfloat[]{\includegraphics[width=0.32\textwidth]{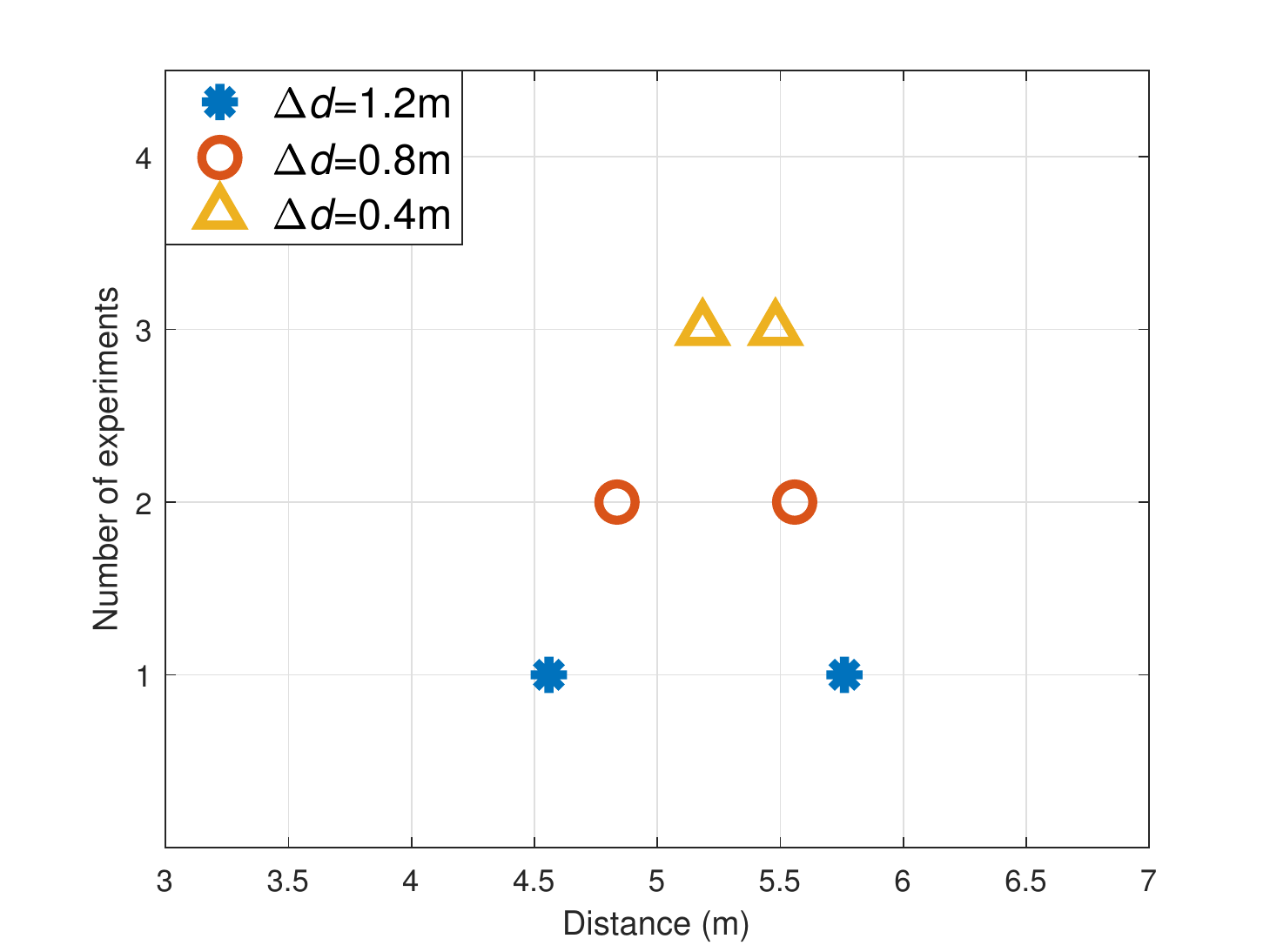} 
		\label{figESPRIT_TAO}}
        \caption{Experiment results with varying distance difference by using Periodogram \protect\subref{figPER_TAO}, MUSIC \protect\subref{figMUSIC_TAO} and ESPRIT \protect\subref{figESPRIT_TAO}.}
        \label{fig_TAO}
        \subfloat[]{\includegraphics[width=0.32\textwidth]{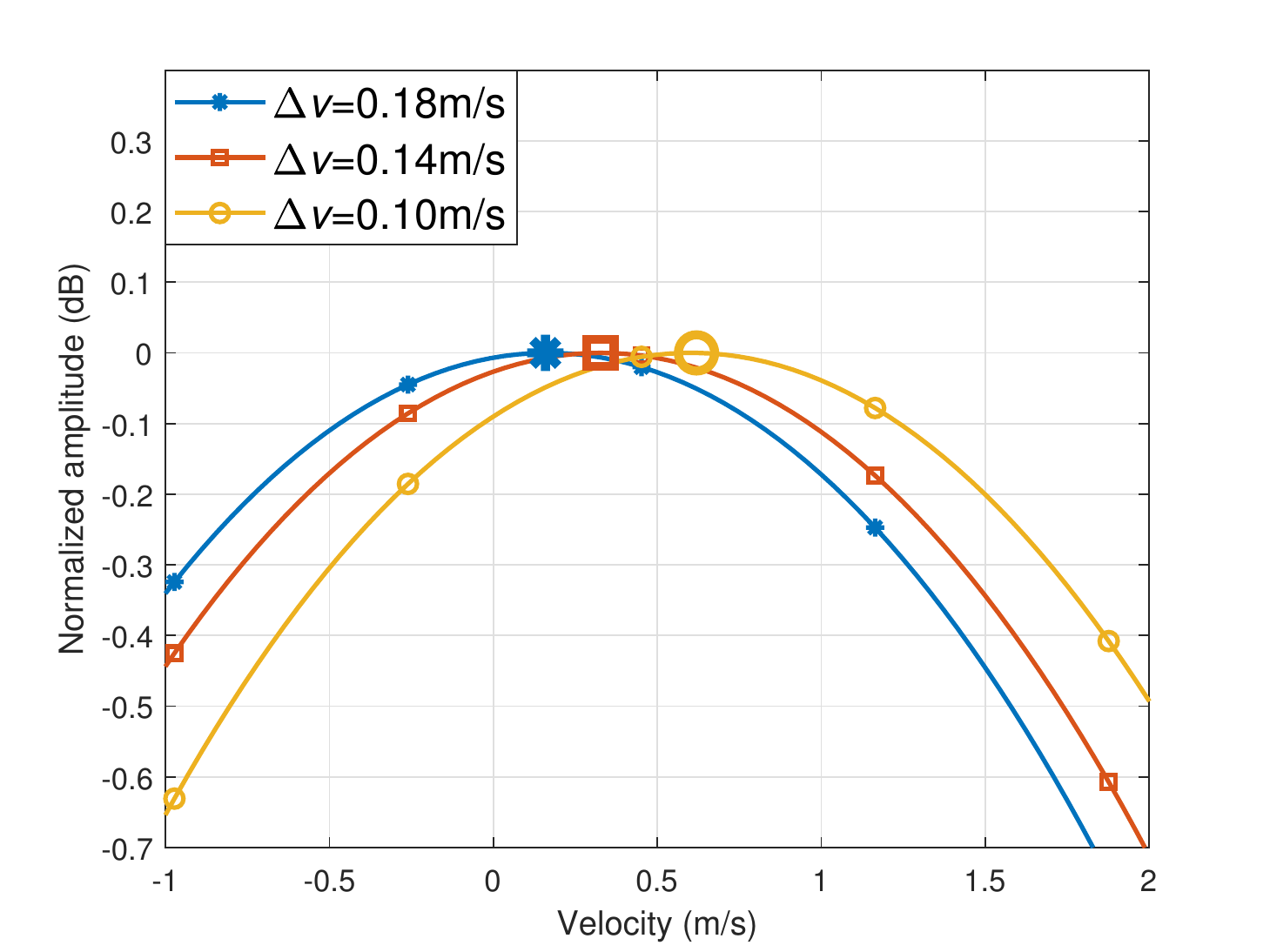}
		\label{figPER_DOP}}
	\hfil
	\subfloat[]{\includegraphics[width=0.32\textwidth]{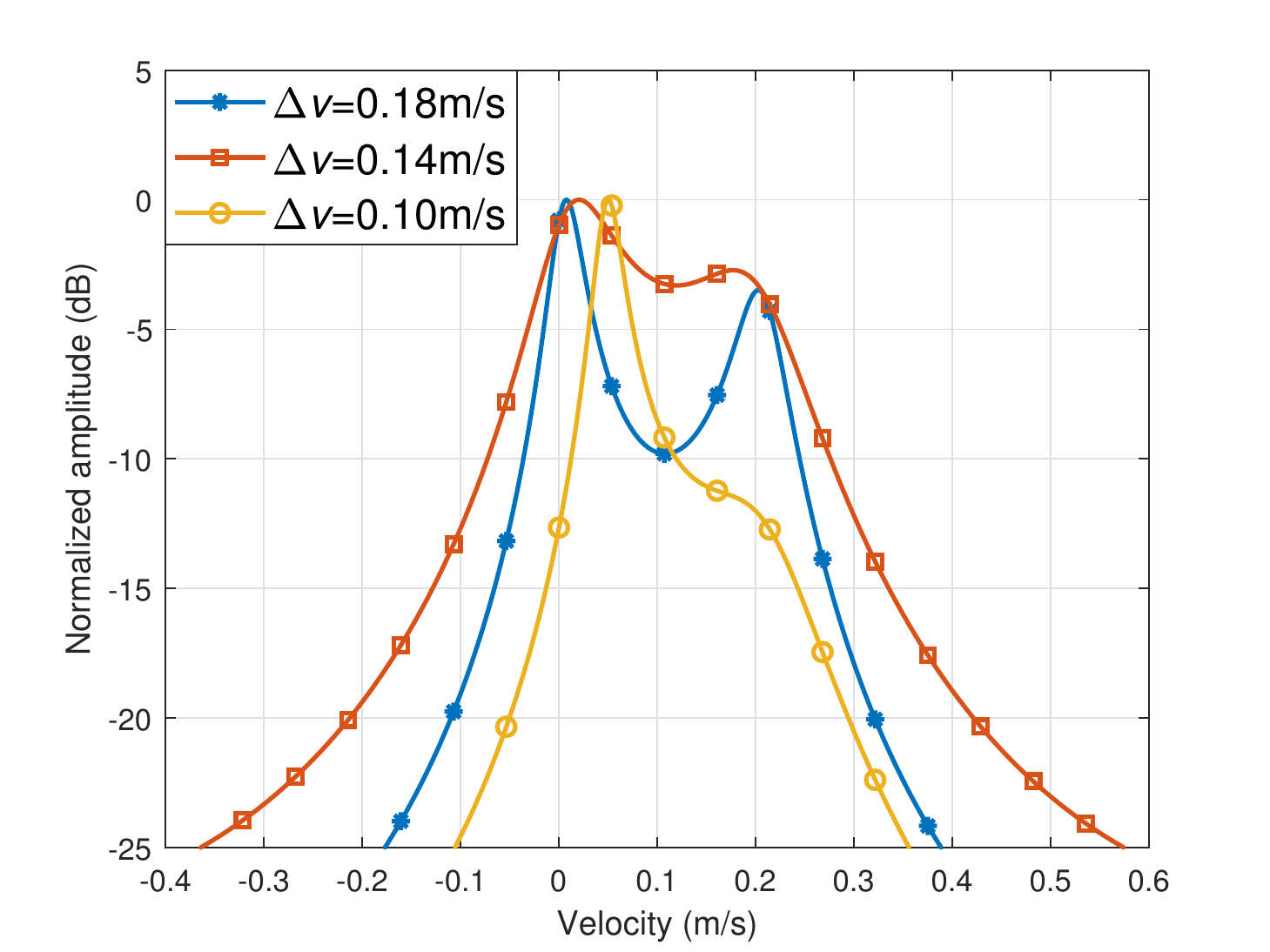}
		\label{figMUSIC_DOP}}
        \hfil
	\subfloat[]{\includegraphics[width=0.32\textwidth]{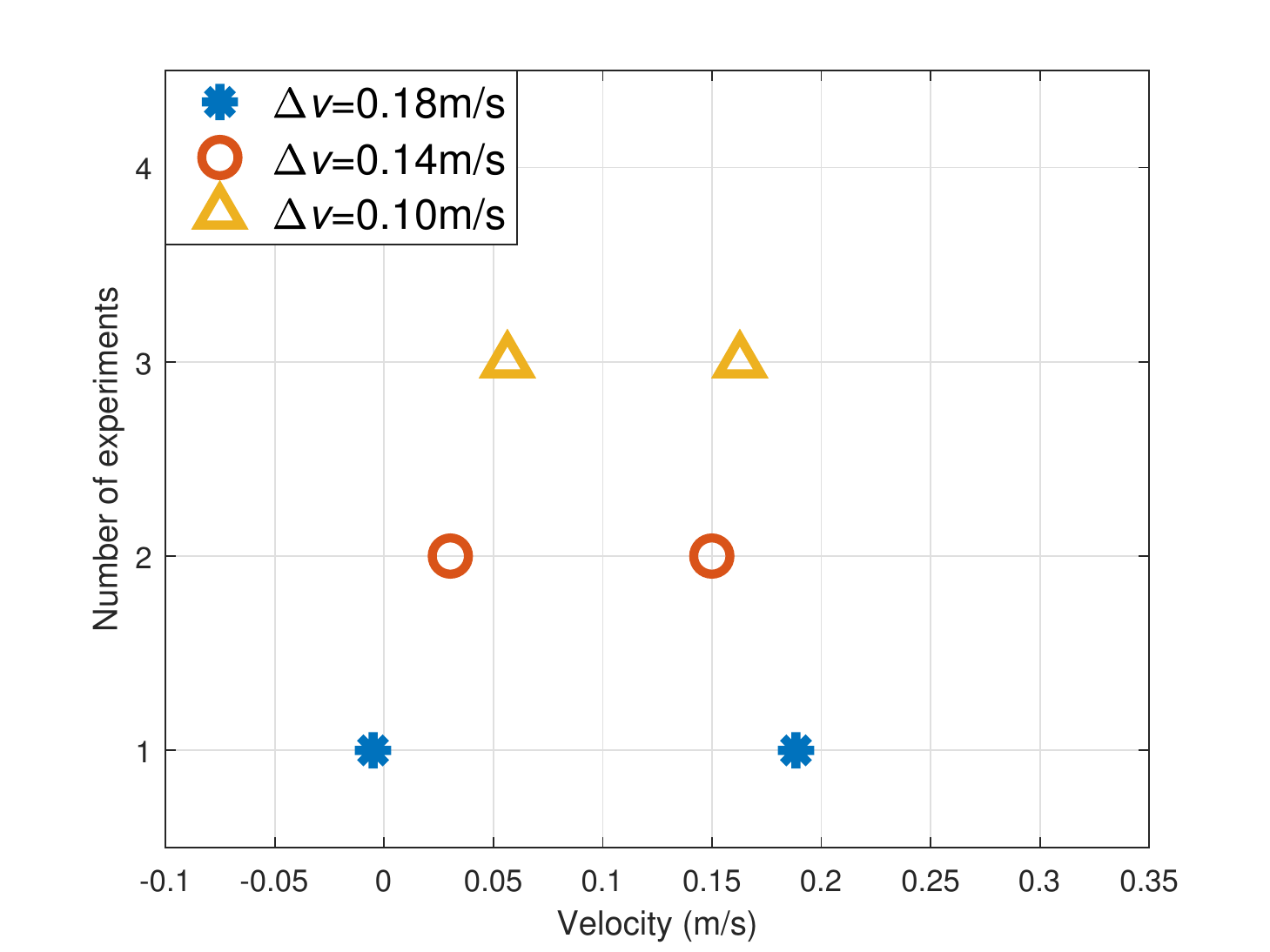} 
		\label{figESPRIT_DOP}}
		\caption{Experiment results with varying velocity difference by using Periodogram \protect\subref{figPER_DOP}, MUSIC \protect\subref{figMUSIC_DOP} and ESPRIT \protect\subref{figESPRIT_DOP}.}
	\label{fig_DOP}
\end{figure*}

The performance of delay estimation for the three algorithms is shown in Fig. \ref{fig_TAO} by evaluating the resolution in terms of various distance differences between the two targets. Fig. \ref{figPER_TAO} and Fig. \ref{figMUSIC_TAO} plot the search spectra versus distance with varying range differences for Periodogram and MUSIC respectively, while Fig. \ref{figESPRIT_TAO} plots the results of five estimations for ESPRIT.
It can be seen that the MUSIC and ESPRIT algorithms can achieve a more accurate estimation than the Periodogram. 
Furthermore, when the distance difference between targets is small, e.g., \SI{0.4}{\meter}, Periodogram fails to distinguish the two targets, while MUSIC and ESPRIT algorithms can resolve the nearby targets, showing their super-resolution capability.
\begin{table}[h]
\center
\caption{Distance of targets}
\begin{tabular}{|c|c|c|c|}
\hline
\textbf{Experiment} & \textbf{Distance 1} & \textbf{Distance 2} & \textbf{Distance difference, $\Delta d$} \\ \hline
1                   & \SI{4.5}{\meter}              & \SI{5.7}{\meter}              & \SI{1.2}{\meter}                 \\ \hline
2                   & \SI{4.7}{\meter}              & \SI{5.5}{\meter}              & \SI{0.8}{\meter}                 \\ \hline
3                   & \SI{4.9}{\meter}              & \SI{5.3}{\meter}              & \SI{0.4}{\meter}                 \\ \hline
\end{tabular}
\label{Distance}
\end{table}

\subsection{Doppler Estimation}
To illustrate velocity estimation, the sliding rails are controlled to move the targets at different speeds, as shown in Table \ref{velocity}.
Limited by hardware conditions, the maximum speed of the sliding rail is \SI[per-mode=symbol]{0.18}{\meter\per\second}.
Similar to the delay estimation experiment, the BS sends a downlink NR frame to initiate the measurements. 
After completing the measurements, the BS uses Periodogram, the method 3 of MUSIC and the method 2 of ESPRIT to process the data, respectively.

Fig. \ref{fig_DOP} shows the performance of the three algorithms in terms of the difference between the velocity of the two targets. 

It is observed that for the Periodogram algorithm, the velocity resolution is so poor that none of the targets can be distinguished at such low speed. 
By contrast, the MUSIC and ESPRIT algorithms can resolve the two targets with small velocity difference, except when the difference is too small, e.g., \SI[per-mode=symbol]{0.10}{\meter\per\second}.

\begin{table}[h]
\center
\caption{Velocity of targets}
\begin{tabular}{|c|c|c|c|}
\hline
\textbf{Experiment} & \textbf{Velocity 1} & \textbf{Velocity 2} & \textbf{Velocity difference, $\Delta v$} \\ \hline
1  & 0 & \SI[per-mode=symbol]{0.18}{\meter\per\second}  & \SI[per-mode=symbol]{0.18}{\meter\per\second}                 \\ \hline
2  & \SI[per-mode=symbol]{0.02}{\meter\per\second} & \SI[per-mode=symbol]{0.16}{\meter\per\second}  & \SI[per-mode=symbol]{0.14}{\meter\per\second}                 \\ \hline
3  & \SI[per-mode=symbol]{0.04}{\meter\per\second} & \SI[per-mode=symbol]{0.14}{\meter\per\second}  & \SI[per-mode=symbol]{0.10}{\meter\per\second}                 \\ \hline
\end{tabular}
\label{velocity}
\end{table}


\section{Conclusion}\label{5 conclu}
In this paper, we propose the concept of ISSAC based on 5G NR waveform, and study the signal processing methods with uneven CPs, including Periodogram, MUSIC and ESPRIT. 
The impact of uneven CPs on various sensing algorithms for Doppler estimation is studied, and three methods are investigated to resolve the issue.  
Finally, an ISSAC experimental platform is built, and two targets with different distance and speed differences are sensed by the ISSAC BS. 
The results show that the two super-resolution algorithms, namely MUSIC and ESPRIT, can significantly improve the delay and Doppler resolution as compared to Periodogram algorithm.

\section*{Acknowledgement}
This work was supported by the National Key R\&D Program of China with Grant number 2019YFB1803400 and by the Natural Science Foundation of China under Grant 62071114.

\bibliographystyle{IEEEtran}
\bibliography{IEEEabrv,reference}

\begin{thebibliography}{10}
\providecommand{\url}[1]{#1}
\csname url@samestyle\endcsname
\providecommand{\newblock}{\relax}
\providecommand{\bibinfo}[2]{#2}
\providecommand{\BIBentrySTDinterwordspacing}{\spaceskip=0pt\relax}
\providecommand{\BIBentryALTinterwordstretchfactor}{4}
\providecommand{\BIBentryALTinterwordspacing}{\spaceskip=\fontdimen2\font plus
\BIBentryALTinterwordstretchfactor\fontdimen3\font minus
  \fontdimen4\font\relax}
\providecommand{\BIBforeignlanguage}[2]{{%
\expandafter\ifx\csname l@#1\endcsname\relax
\typeout{** WARNING: IEEEtran.bst: No hyphenation pattern has been}%
\typeout{** loaded for the language `#1'. Using the pattern for}%
\typeout{** the default language instead.}%
\else
\language=\csname l@#1\endcsname
\fi
#2}}
\providecommand{\BIBdecl}{\relax}
\BIBdecl

\bibitem{liu2022integrated}
F.~Liu, Y.~Cui, C.~Masouros, J.~Xu, T.~X. Han, Y.~C. Eldar, and S.~Buzzi,
  ``Integrated sensing and communications: Towards dual-functional wireless
  networks for 6g and beyond,'' \emph{IEEE J. Sel. Areas Commun.}, 2022.

\bibitem{b59}
F.~Liu, C.~Masouros, T.~Ratnarajah, and A.~Petropulu, ``On range sidelobe
  reduction for dual-functional radar-communication waveforms,'' \emph{{IEEE}
  Wireless Commun. Lett.}, vol.~9, no.~9, pp. 1572--1576, 2020.

\bibitem{waveform}
Z.~Xiao and Y.~Zeng, ``Waveform design and performance analysis for full-duplex
  integrated sensing and communication,'' \emph{IEEE J. Sel. Areas Commun.},
  vol.~40, no.~6, pp. 1823--1837, 2022.

\bibitem{coverage}
R.~Li, Z.~Xiao, and Y.~Zeng, ``Beamforming towards seamless sensing coverage
  for cellular integrated sensing and communication,'' in \emph{Proc. IEEE Int.
  Conf. Commun. Workshops (ICC Workshops)}.\hskip 1em plus 0.5em minus
  0.4em\relax IEEE, 2022, pp. 492--497.

\bibitem{hua2023optimal}
H.~Hua, J.~Xu, and T.~X. Han, ``Optimal transmit beamforming for integrated
  sensing and communication,'' \emph{IEEE Transactions on Vehicular
  Technology}, 2023.

\bibitem{b74}
S.~Chen, Z.~Xiao, and Y.~Zeng, ``Simultaneous beam sweeping for multi-beam
  integrated sensing and communication,'' in \emph{Proc. IEEE Int. Conf.
  Commun. (ICC)}, 2022, pp. 4438--4443.

\bibitem{b76}
I.~S. Gradshteyn and I.~M. Ryzhik, \emph{Table of integrals, series, and
  products}.\hskip 1em plus 0.5em minus 0.4em\relax Academic press, 2014.

\bibitem{b64}
M.~Kobayashi, G.~Caire, and G.~Kramer, ``Joint state sensing and communication:
  Optimal tradeoff for a memoryless case,'' in \emph{Proc. IEEE Int. Symp. Inf.
  Theor. Proc.}, 2018, pp. 111--115.

\bibitem{crb}
H.~Wang, Z.~Xiao, and Y.~Zeng, ``Cram\'{e}r-rao bounds for near-field sensing
  with extremely large-scale {MIMO},'' \emph{arXiv preprint arXiv:2303.05736},
  2023.

\bibitem{snr}
H.~Wang and Y.~Zeng, ``{SNR} scaling laws for radio sensing with extremely
  large-scale {MIMO},'' in \emph{Proc. IEEE Int. Conf. Commun. Workshops (ICC
  Workshops)}, 2022, pp. 121--126.

\bibitem{sturm2011gemeinsame}
C.~Sturm, ``Joint implementation of communication and radar based on {OFDM},''
  Ph.D. dissertation, Karlsruhe Institute of Technology, 2011.

\bibitem{2014ofdmradar}
K.~M. Braun, ``{OFDM} radar algorithms in mobile communication networks,''
  Ph.D. dissertation, Karlsruhe, Karlsruher Institut f{\"u}r Technologie (KIT),
  Diss., 2014.

\bibitem{li2019conditioning}
W.~Li and W.~Liao, ``Conditioning of restricted fourier matrices and
  super-resolution of {MUSIC},'' in \emph{Int. Conf. Sampl. Theory Appl.,
  SampTA}.\hskip 1em plus 0.5em minus 0.4em\relax IEEE, 2019, pp. 1--4.

\bibitem{b46}
R.~Roy and T.~Kailath, ``{ESPRIT}-estimation of signal parameters via
  rotational invariance techniques,'' \emph{{IEEE} Trans. Acoust., Speech,
  Signal Process.}, vol.~37, no.~7, pp. 984--995, 1989.

\bibitem{gonen1999subspace}
E.~Gonen and J.~M. Mendel, ``Subspace-based direction finding methods,''
  \emph{The digital signal processing handbook}, vol.~62, 1999.

\bibitem{radarp}
P.~Peebles, Jr., \emph{Radar Principles}.\hskip 1em plus 0.5em minus
  0.4em\relax John Wiley and Sons Inc., 1998.

\bibitem{delay_and_Doppler}
D.~H. Nguyen and R.~W. Heath, ``Delay and doppler processing for multi-target
  detection with ieee 802.11 {OFDM} signaling,'' in \emph{2017 ICASSP IEEE Int.
  Conf. Acoust. Speech Signal Process Proc.}\hskip 1em plus 0.5em minus
  0.4em\relax IEEE, pp. 3414--3418.

\bibitem{zheng2017super}
L.~Zheng and X.~Wang, ``Super-resolution delay-doppler estimation for {OFDM}
  passive radar,'' \emph{{IEEE} Trans. Signal Process.}, vol.~65, no.~9, pp.
  2197--2210, 2017.

\bibitem{xie2021performance}
R.~Xie, D.~Hu, K.~Luo, and T.~Jiang, ``Performance analysis of joint
  range-velocity estimator with 2d-music in {OFDM} radar,'' \emph{{IEEE} Trans.
  Signal Process.}, vol.~69, pp. 4787--4800, 2021.

\bibitem{isac5g}
Z.~Wei, H.~Qu, Y.~Wang, X.~Yuan, H.~Wu, Y.~Du, K.~Han, N.~Zhang, and Z.~Feng,
  ``Integrated sensing and communication signals towards {5G-A} and {6G}: A
  survey,'' \emph{{IEEE} Internet Things J.}, pp. 1--1, 2023.

\bibitem{henninger2022computationally}
M.~Henninger, S.~Mandelli, M.~Arnold, and S.~Ten~Brink, ``A computationally
  efficient {2D} music approach for {5G} and {6G }sensing networks,'' in
  \emph{2022 IEEE Wireless Commun. Networking Conf. WCNC}.\hskip 1em plus 0.5em
  minus 0.4em\relax IEEE, 2022, pp. 210--215.

\bibitem{pucci2021performance}
L.~Pucci, E.~Matricardi, E.~Paolini, W.~Xu, and A.~Giorgetti, ``Performance
  analysis of joint sensing and communication based on {5G} new radio,'' in
  \emph{2021 IEEE Globecom Workshops (GC Wkshps)}.\hskip 1em plus 0.5em minus
  0.4em\relax IEEE, 2021, pp. 1--6.

\bibitem{3gpp38211}
3GPP, ``{NR; Physical channels and modulation },'' {3rd Generation Partnership
  Project (3GPP)}, Technical Specification (TS) 38.211, 01 2023, version
  17.4.0.

\bibitem{An_Experimental_Proof}
T.~Xu, F.~Liu, C.~Masouros, and I.~Darwazeh, ``An experimental proof of concept
  for integrated sensing and communications waveform design,'' \emph{IEEE open
  J. Commun. Soc.}, vol.~3, pp. 1643--1655, 2022.

\bibitem{An_Experimental_Study}
M.~Temiz, C.~Horne, N.~J. Peters, M.~A. Ritchie, and C.~Masouros, ``An
  experimental study of radar-centric transmission for integrated sensing and
  communications,'' \emph{{IEEE} Trans. Microw. Theory Techn.}, pp. 1--14,
  2023.

\bibitem{First_Demonstration}
J.~Wang, X.-D. Liang, L.-Y. Chen, L.-N. Wang, and K.~Li, ``First demonstration
  of joint wireless communication and high-resolution sar imaging using
  airborne {MIMO} radar system,'' \emph{{IEEE} Trans. Geosci. Remote Sens.},
  vol.~57, no.~9, pp. 6619--6632, 2019.

\bibitem{Joint_Communication}
Q.~Zhang and X.~Gao, ``Joint communication and sensing enabled cooperative
  perception testbed for connected automated vehicles,'' in \emph{INFOCOM
  WKSHPS 2022 - IEEE Conf. Comput. Commun. Workshops}, 2022, pp. 1--2.

\bibitem{Spatial_Modulation}
D.~Ma, N.~Shlezinger, T.~Huang, Y.~Shavit, M.~Namer, Y.~Liu, and Y.~C. Eldar,
  ``Spatial modulation for joint radar-communications systems: Design,
  analysis, and hardware prototype,'' \emph{{IEEE} Trans. Veh. Technol.},
  vol.~70, no.~3, pp. 2283--2298, 2021.

\bibitem{Multifunctional_Transceiver}
J.~Moghaddasi and K.~Wu, ``Multifunctional transceiver for future radar sensing
  and radio communicating data-fusion platform,'' \emph{{IEEE} Access}, vol.~4,
  pp. 818--838, 2016.

\bibitem{3gpp38201}
3GPP, ``{NR; Physical layer; General description},'' {3rd Generation
  Partnership Project (3GPP)}, Technical Specification (TS) 38.201, 05 2022,
  version 17.0.0.

\bibitem{b73}
C.~Sturm and W.~Wiesbeck, ``Waveform design and signal processing aspects for
  fusion of wireless communications and radar sensing,'' \emph{Proc. IEEE},
  vol.~99, no.~7, pp. 1236--1259, 2011.

\bibitem{2018VNC}
C.~D. Ozkaptan, E.~Ekici, O.~Altintas, and C.-H. Wang, ``{OFDM} pilot-based
  radar for joint vehicular communication and radar systems,'' in \emph{2018
  IEEE Veh. Netw. Conf., VNC}, 2018, pp. 1--8.

\bibitem{Superresolution_techniques_for_time_domain}
H.~Yamada, M.~Ohmiya, Y.~Ogawa, and K.~Itoh, ``Superresolution techniques for
  time-domain measurements with a network analyzer,'' \emph{{IEEE} Trans.
  Antennas Propag.}, vol.~39, no.~2, pp. 177--183, 1991.

\end{thebibliography}

\end{document}